\documentclass[preprint,aip,jcp]{revtex4-1}
\usepackage[utf8]{inputenc}

\usepackage{epsfig}
\usepackage{textcomp}
\usepackage{graphicx}
\usepackage{amsmath}
\usepackage{amssymb}
\usepackage{amsmath}
\usepackage{url}
\usepackage[normalem]{ulem}
\renewcommand{\d}{\text{d}}
\usepackage{color}
\definecolor{Myblue}{rgb}{0.3,0.3,1.0}
\usepackage{hyperref}

\csname @ifundefined\endcsname{endmcitethebibliography}
{}{}

\usepackage{rotating}

\newcommand{\OA}{O$_{\rm A}$}
\newcommand{\OB}{O$_{\rm B}$}
\newcommand{\HA}{H$_{\rm A}$}
\newcommand{\HB}{H$_{\rm B}$}

\begin{document}
\title{OH-Formation Following Vibrationally Induced Reaction Dynamics
  of H$_2$COO}

\author{Kaisheng Song}\altaffiliation{School of Chemistry and Chemical
  Engineering, Chongqing University, Chongqing 401331, China}
\affiliation{Department of Chemistry, University of Basel,
  Klingelbergstrasse 80, CH-4056 Basel, Switzerland}
  
\author{Meenu Upadhyay} \affiliation{Department of Chemistry,
  University of Basel, Klingelbergstrasse 80, CH-4056 Basel,
  Switzerland}

\author{Markus Meuwly} \email[]{m.meuwly@unibas.ch}
\affiliation{Department of Chemistry, University of Basel,
  Klingelbergstrasse 80, CH-4056 Basel, Switzerland}

\date{\today}

\begin{abstract}
The reaction dynamics of H$_2$COO to form linear HCOOH and dioxirane
as first steps for OH-elimination is quantitatively
investigated. Using a machine learned potential energy surface at the
CASPT2/aug-cc-pVTZ level of theory vibrational excitation along the
CH-normal mode $\nu_{\rm CH}$ with energies up to 40.0 kcal/mol ($\sim
5 \nu_{\rm CH}$) leads almost exclusively to linear HCOOH which
further decomposes into OH+HCO. Although the barrier to form dioxirane
is only 21.4 kcal/mol the reaction probability to form dioxirane is
two orders of magnitude lower if the CH-stretch mode is
excited. Following the dioxirane-formation pathway is facile, however,
if in addition the COO-bend vibration is excited with energies
equivalent to $\sim (2 \nu_{\rm CH} + 4 \nu_{\rm COO})$ or $\sim (3
\nu_{\rm CH} + \nu_{\rm COO})$. For OH-formation in the atmosphere the
pathway through linear HCOOH is probably most relevant because the
alternative pathways (through dioxirane or formic acid) involve
several intermediates that can de-excite through collisions, relax
{\it via} Intramolecular vibrational energy redistribution (IVR), or
pass through very loose and vulnerable transition states (formic
acid). This work demonstrates how, by selectively exciting particular
vibrational modes, it is possible to dial into desired reaction
channels with a high degree of specificity for a process relevant to
atmospheric chemistry.
\end{abstract}

\maketitle

\section{Introduction}
The photodissociation dynamics of small molecules is of fundamental
interest in atmospheric chemistry. One of the chemically most relevant
agents is the hydroxyl radical (OH)\cite{stone:2012} which was also
referred to as the ``detergent of the
troposphere''.\cite{gligorovski2015environmental,levy1971normal} The
radical triggers degradation of pollutants including volatile organic
compounds (VOCs) and is an important chain initiator in most oxidation
processes in the atmosphere. The amount of OH generated from alkene
ozonolysis is an important determinant required for chemical models of
the lower atmosphere. Field studies have suggested that ozonolysis of
alkenes is responsible for the production of about one third of the
atmospheric OH radicals during daytime, and is the predominant source
of OH radicals at night.\cite{emmerson2009night,khan2018criegee}
Alkene ozonolysis proceeds through a 1,3-cycloaddition of ozone across
the C=C bond to form a primary ozonide which then decomposes into
carbonyl compounds and energized carbonyl oxides, known as Criegee
Intermediates (CIs).\cite{criegee1949ozonisierung} These energized
intermediates rapidly undergo either unimolecular decay to hydroxyl
radicals\cite{alam2011total} or collisional
stabilization.\cite{novelli2014direct} Stabilized CIs can isomerize
and decompose into products including the OH radical, or undergo
bimolecular reactions with water vapor, SO$_2$, NO$_2$ and
acids\cite{taatjes2017criegee,mauldin2012new}.\\

\begin{figure}
    \centering
    \includegraphics[width=\textwidth]{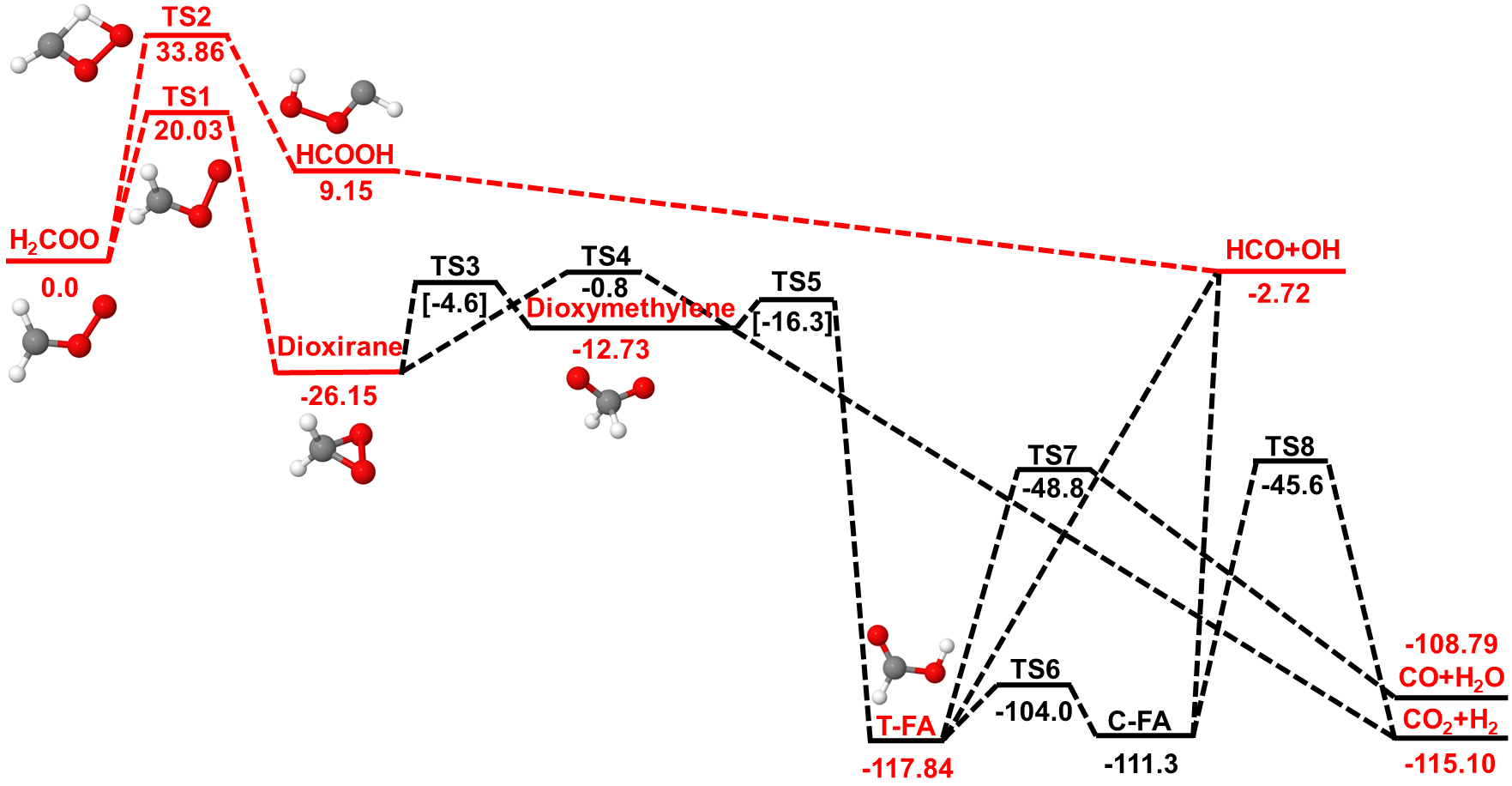}
    \caption{Decomposition pathways for H$_2$COO on the singlet PES at
      the CCSD(T)-F12a/aVTZ level of theory (red). Two reaction
      channels (red line) are considered in the present work: The
      H-transfer channel leading to linear HCOOH which further decays
      to HCO+OH and the dioxirane channel which leads to formic acid
      (FA) and three different final product channels (the
      H-production channels are omitted for clarity).  Here, the
      reaction channels (black line, energies in black determined at
      CCSD(T)/aVTZ//B3LYP/aVTZ level, and the energies in the square
      bracket determined at CCSD(T)/aVTZ//CASSCF(8,8)/cc-pVTZ level)
      were those from Ref.\cite{nguyen:2015}. The present work
      investigates the first reactive step along the two
      pathways. Barriers for formation of the CO+H$_2$O and
      CO$_2$+H$_2$ are $\sim 70$ kcal/mol. A third possible pathway
      involving oxygen-atom insertion into one of the CH-bonds to
      yield formic acid is not shown in this chart.\cite{nguyen:2015}}
    \label{fig:pathway}
\end{figure}

\noindent
The smallest CI is formaldehyde oxide (H$_2$COO). Laboratory studies
required for a more detailed understanding of the spectroscopy and
reaction dynamics\cite{lee:2015} became possible following successful
in situ generation of H$_2$COO using photolysis of CH$_2$I$_2$ in
O$_2$.\cite{welz:2012} Earlier computations proposed that energized
H$_2$COO can decompose to HCO+OH and H$_2$CO+O$(^3 {\rm P})$ or
isomerize to dioxirane. Access to the HCO+OH channel requires
H-transfer to form the linear isomer HCOOH. The dioxirane and
H-transfer pathways are shown in Figure \ref{fig:pathway}. \\

\noindent
Vibrationally induced reactivity has been found to initiate a sequence
of chemical transformations in the next-larger CI, {\it
  syn-}CH$_3$CHOO. Direct time-domain experimental rates for
appearance of OH from unimolecular dissociation of
\textit{syn}-CH$_3$CHOO under collision free conditions were obtained
by vibrationally activating the molecules with energies equivalent to
approximately two quanta in the CH-stretch
vibration\cite{fang:2016,fang:2016deep} which is close to the barrier
for formation of H$_2$CCHOOH and subsequent
OH-elimination. Computationally, the entire reaction pathway from
energized {\it syn-}CH$_3$CHOO to OH(X$^2 \Pi$) elimination was
followed using neural network (NN) representations of the potential
energy surfaces (PESs).\cite{MM.criegee:2021} In addition to
OH-elimination, OH-roaming and formation of glycolaldehyde was found
as an alternative reaction pathway.\cite{MM.criegee:2023}\\

\noindent
Excitation of internal vibrational modes was also proposed as a means
for OH-elimination in species relevant to atmosphere chemistry,
including HONO, HONO$_2$, or HO$_2$NO$_2$. These species can absorb
visible radiation and induce vibrational overtone transitions to
states with several quanta in the OH stretching
vibration.\cite{Donaldson:1997} For H$_2$SO$_4$ (sulfuric acid)
vibrationally induced reactivity by exciting the OH-stretch mode was
implicated in photodissociation dynamics forming SO$_3$ and
water.\cite{Vaida:2003} Subsequent molecular dynamics simulations at
various levels of sophistication confirmed that excitation of the
OH-stretch with 4 to 5 quanta drives the decomposition of
H$_2$SO$_4$.\cite{Miller:2006,yosa:2011,reyes.pccp.2014.msarmd}
However, although cavity ring-down spectroscopy successfully probed
the asymmetric OH stretching vibration with $\nu_9 = 4$ and $\nu_9 =
5$, vibrationally induced photodissociation dynamics has as yet not
been observed directly.\cite{Feierabend:2006}\\

\noindent
In the present work, the reaction dynamics of the smallest CI,
H$_2$COO, following vibrational excitation of internal vibrational
modes is considered in order to characterize two competing, low-energy
pathways. Excitation of the CH-stretch mode was demonstrated to
initiate chemical processing for {\it syn-}CH$_3$CHOO both, from
experiments\cite{liu:2014} and atomistic
simulations.\cite{MM.criegee:2021,MM.criegee:2023} For H$_2$COO
oxygen-atom elimination requires up to 50 kcal/mol whereas the
H-transfer and dioxirane channels feature approximate barrier heights
of 34 and 20 kcal/mol, respectively. Following the H-transfer pathway
directly yields OH+HCO whereas the dioxirane pathway leads to formic
acid which may stabilize through collisions or follow further chemical
processing, see Figure \ref{fig:pathway}. First, the methods are
presented, followed by a description of the intermolecular
interactions and the dynamics and rates of the photodissociation
reaction. At the end, conclusions are drawn.\\

\section{Methods}
In this section the reference electronic structure calculations and
the construction of the reactive, multidimensional PESs is
described. Two representations are considered in the following: a
neural network-based approach using the PhysNet architecture and a
more empirical but computationally considerably more efficient
multi-state adiabatic reactive MD (MS-ARMD) representation. One
important difference between these two representations is that MS-ARMD
requires a dedicated model to follow one reaction pathway at a time
whereas the NN-based representation is capable of describing the
dynamics along both pathways at the same time.\\

\subsection{Electronic Structure Calculations}
Reference energies were first determined at the explicitly correlated
(F12) coupled cluster method with singles, doubles, and perturbative
triples\cite{adler2007F12,knizia2009ccsd(t)-f12} with the augmented
correlation-consistent polarized valence triple-$\zeta$ basis set
(CCSD(T)-F12a/aVTZ). For transfer learning, {\it vide infra},
additional CASPT2/aVTZ calculations were carried out. All electronic
structure calculations used the Molpro2019 suite of
codes.\cite{MOLPRO2019}\\

\noindent
Initially, $\sim 6000$ structures for each H$_2$COO and Dioxirane were
selected from two existing datasets that were used to construct two
PIP-NN PESs from reference calculations at the CCSD(T)-F12a/aVTZ
level.\cite{Jun2014H2COO,Jun2016dioxirane} Additional reference
structures were generated around the two TSs and all the minima
considered in the present work by scanning a regular grid in internal
coordinates. Subsequently, the "base model" was trained (see below)
using energies and forces at this level of theory. The ``base model''
was further improved from several rounds of adaptive sampling. This
lead to a total of 29612 structures covering a wide configurational
space covering the H-transfer and dioxirane formation channels.\\

\noindent
In order to explore whether alternative feasible reaction channels at
the conditions considered in the dynamics simulations exist,
one-dimensional rigid scan at the CASPT2 level of theory along the
O--O bond for the reactant H$_2$COO at its minimum energy structure is
shown in Figure \ref{fig:scan-oo-ch}. The potential energy curve along
the O--O bond agrees well with existing results at the MRCI-F12 level
of theory,\cite{dawes2015uv} where the O--O separation is 1.34 \AA\/
at equilibrium, and the well depth is $\sim 50.0$ kcal/mol. The
barriers for O-O dissociation and C-H dissociation with 55.00 and
131.00 kcal/mol are significantly higher than the barriers for
H-transfer (35.47 kcal/mol) and dioxirane formation (21.39
kcal/mol). Hence, for exploring the H-transfer and dioxirane channels,
there is no need to include reference structures for the O + CH$_2$O
and H + CHOO channels in the present work. The energy profiles of both
reaction channels from different computational levels are shown in
Figure \ref{fig:mep}. Both methods CCSD(T)-F12a and CASPT2 are
reliable for developing PESs for the title reaction in this
work. Conversely, noticeable underestimation is observed in the
results obtained from MRCI(12,11) calculations for the dioxirane
formation channel subsequent to the transition state. Hence, MRCI is
not the best candidate in this work.  \\

\subsection{Training of the Neural Network and Transfer Learning}
Machine-learned PESs were trained based on the PhysNet
architecture,\cite{MM.physnet:2019} which is designed for predicting
energies, forces, dipole moments, and partial charges. The technical
background for PhysNet has been detailed in
Ref.\cite{MM.physnet:2019}. By construction, the PhysNet model is
permutationally invariant. For the base model, energies, forces and
dipole moments for 29612 structures were determined at the
CCSD(T)-F12a/aVTZ level of theory. Four independent PhysNet models
were trained, using 80\% of the data as the training set whereas test
and validation sets each contained 10\% of the data. The performance
on the test data is reported in Figure \ref{fig:corr-ccsdt}. The best
of the four base models was selected for subsequent simulations.\\

\noindent
Some of the structural rearrangements are likely to require
multi-reference descriptions of the electronic structure. To this end,
transfer learning from the CCSD(T)/aVTZ to the CASPT2/aVTZ level of
theory was used to improve the PES in regions where multi reference
effects may become relevant, for example for O--O bond breaking after
H-transfer.  Transfer learning (TL) has been shown to be a valuable
and resource-efficient technique for developing global potential
energy surfaces starting from models based on different initial
calculations.\cite{MM.tl:2022,MM.tl:2023,MM.nnpes:2023,kaser2022transfer,smith:2019}
Here, energies and forces for 2000 structures along the IRCs were used
together with several hundred structures around the minima and
transition states for both pathways (Figure \ref{fig:pathway}) from
which a preliminary TL PES was trained. Next, adaptive sampling was
employed to refine the dataset for the TL PES. For this, DMC and short
MD simulations with 40.0 kcal/mol excess energy along the CH-stretch
mode were run to validate the TL PES and identify deficiencies. The
final data set contained 5162 structures for which energies and forces
at the CASPT2/aug-cc-pVTZ level were used to train 4 independent
models. For TL, the data was split into training (90\%), validation
(5\%), and test set (5\%). After that, the best of the four models was
chosen for production simulations.\\

\subsection{Fitting the MS-ARMD PES}
MS-ARMD is a computationally efficient means to investigate chemical
reactions based on empirical force fields.\cite{MM.armd:2014} The
initial parameters for the reactant (H$_2$COO) and products (HCOOH and
dioxirane) were taken from SwissParam \cite{zoete2011swissparam}.
First, the representative structures were sampled from 500 ps MD
simulations at 1000 K, and energies were determined at
CCSD(T)-F12A/aug-cc-pVTZ level of theory. The force fields for
reactant and products were separately parametrized using a downhill
simplex algorithm \cite{nelder1965simplex}. Several rounds of
parameter refinements were done until the root mean squared deviation
for the final set between the target (\textit{ab initio}) and the
fitted energies for H$_2$COO, HCOOH and dioxirane reached 0.85/1.5 and
1.07 and 1.7 kcal/mol, respectively. GAPO functions
\cite{MM.armd:2014} were used to connect reactant and product force
fields to yield a continuous, reactive PES along the reaction
paths. The transition states for hydrogen transfer and dioxirane
formation were 33.83 and 20.33 kcal/mol, respectively, which compare
well with reference energies of 33.86 and 20.03 kcal/mol. All
parameters for the MS-ARMD PESs are given in the supporting
information, see Tables \ref{tab:dioxirane_channel} to
\ref{tab:gapo-htransfer}.  \\

\begin{figure}
    \centering \includegraphics[width=1.0\textwidth]{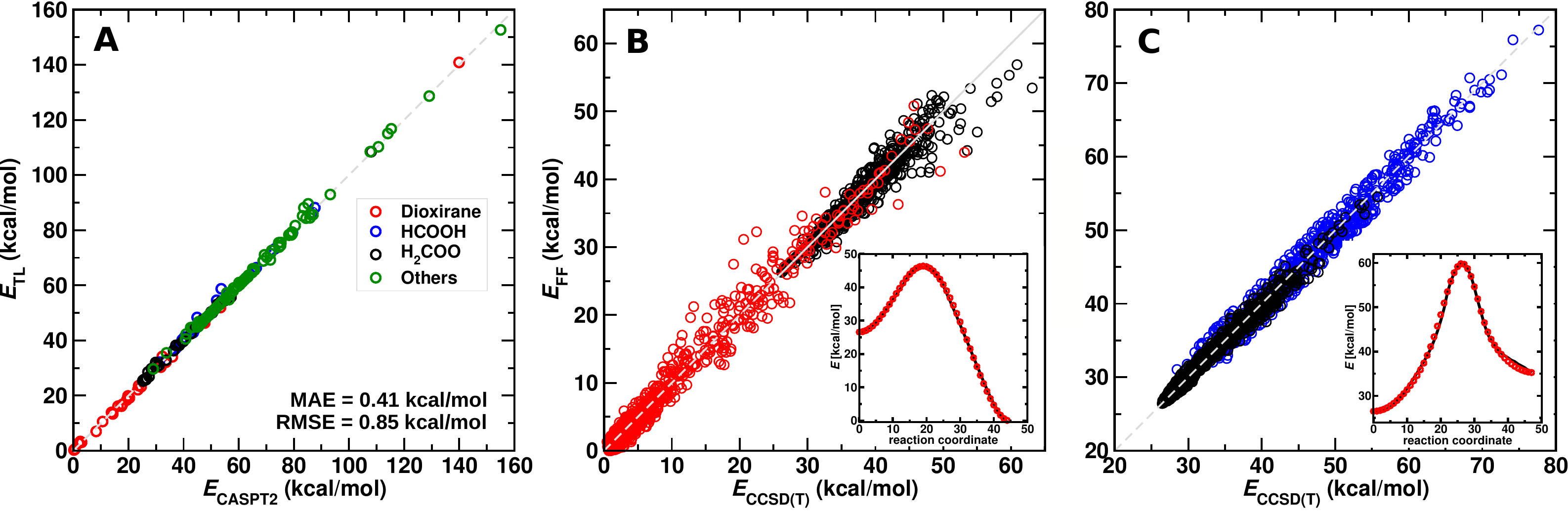}
    \caption{Correlation of 259 (5\%) \textit{ab initio} reference
      energies and predicted TL energies for the test set from the
      PhysNet TL model. Panel B: Correlation of the \textit{ab initio}
      reference structures and the fitted force field for H$_2$COO
      (black) and dioxirane (red) with RMSE values of 1.5 and 1.7
      kcal/mol respectively. Inset: \textit{ab initio} IRC (red
      circles) and fitted MS-ARMD (black curve, after GAPOs
      fitting). Panel C: Correlation of the \textit{ab initio}
      reference structures and the fitted force field for H$_2$COO
      (black) and HCOOH (blue) with RMSE values of 0.85 and 1.07
      kcal/mol respectively. Inset: \textit{ab initio} IRC (red
      circles) and fitted MS-ARMD (black curve, after GAPOs fitting).}
    \label{fig:quality}
\end{figure}

\subsection{Molecular Dynamics Simulations and Analysis}
All reactive molecular dynamics simulations were carried out using the
CHARMM molecular simulation program\cite{charmm:2009} including
provisions for reactive MD (MS-ARMD)
simulations\cite{MM.rmd:2008,MM.armd:2014} and pyCHARMM/PhysNet for
the ML/MD simulations.\cite{MM.pycharmm:2023,pycharmm:2023} All
simulations were run in the \textit{NVE} ensemble and in the gas
phase. The MD simulations were initiated from the minimized structure
of H$_2$COO, using the PhysNet PESs for pyCHARMM simulations and the
MS-ARMD PESs for CHARMM simulations. The time step in all simulations
was $\Delta t = 0.1$ fs to conserve total energy as the bonds
involving hydrogen atoms were flexible and each trajectory was run for
1 ns.\\

\noindent
For the PhysNet TL PES, simulations involving vibrational excitation
were carried out akin to earlier investigations of the
photodissociation of
syn-CH$_3$CHOO.\cite{MM.criegee:2021,MM.criegee:2023}The protocol for
obtaining initial coordinates and velocities was as follows. First,
the system was heated to 300 K for 200 ps, followed by equilibration
during 50 ps, and a 1 ns production simulation in the microcanonical
$(NVE)$ ensemble, from which coordinates and velocities were extracted
at intervals of 100 fs. This protocol was repeated for 10 independent
trajectories. The coordinates and velocities were then utilized as
initial conditions for subsequent extensive simulations conducted at
various excitation energies.\\

\noindent
Following earlier experiments and simulations for
syn-CH$_3$CHOO,\cite{lester:2016,MM.criegee:2021,MM.criegee:2023}
first excitation along the CH-stretch normal mode was considered. For
this, the instantaneous velocity vector was scaled along the normal
mode of the CH-stretch vibration such as to yield the desired
excitation energy. For the H-transfer channel, excitation energies
ranged from 16 to 40 kcal/mol, corresponding to $\sim 2$ to $\sim 5$
quanta along the CH-stretch vibration, for which 1000 ($\sim 2
\nu_{\rm CH}$), 1000 ($\sim 3 \nu_{\rm CH}$), 6000 ($\sim 4 \nu_{\rm
  CH}$) and 3000 ($\sim 5 \nu_{\rm CH}$) independent trajectories were
propagated for 1 ns each. The number of quanta considered for the
excitation was guided by the barrier heights for the H-transfer and
dioxirane channels which are 35.5 and 21.4 kcal/mol, respectively, at
the CASPT2/aVTZ level of theory.\\

\noindent
Because excitation along the CH-stretch vibration leads to very small
numbers of crossings along the dioxirane channel even for the highest
excitation energy (46/3000 crossings with 40.0 kcal/mol excitation for
a barrier height of 21.4 kcal/mol), different vibrational modes were
considered for this pathway. Guided by the minimum dynamic path,
discussed further below, excitation along the combination of the
CH-stretch and the C\OA\OB{} bending mode $(\nu_{\rm COO})$ was
used. In this case, the instantaneous velocity vector was first scaled
such as to excite the CH-stretch vibration with the desired energies
which were 16.0 and 24.0 kcal/mol, respectively, corresponding to
$\sim 2 \nu_{\rm CH}$ and $\sim 3 \nu_{\rm CH}$. Next, the resulting
velocity vector was scaled along the $\nu_{\rm COO}$ normal mode to
reach total excitation energies of 22.0 and 25.5 kcal/mol,
respectively, equivalent to exciting $\sim (2 \nu_{\rm CH} + 4
\nu_{\rm COO})$ and $\sim (3 \nu_{\rm CH} + 1 \nu_{\rm COO})$.\\

\noindent
Using MS-ARMD the reactant was heated to 300 K and equilibrated for 50
ps followed by free $NVE$ dynamics for 1 ns. Again, coordinates and
velocities were saved to obtain 5000 initial conditions for each of
the excitation energies. Vibrational excitation was accomplished
through the same procedure as for the PhysNet simulations described
above. 2000 independent trajectories for each excitation energy were
run for 1 ns each.\\

\begin{figure}[h!]
    \centering
    \includegraphics[width=\textwidth]{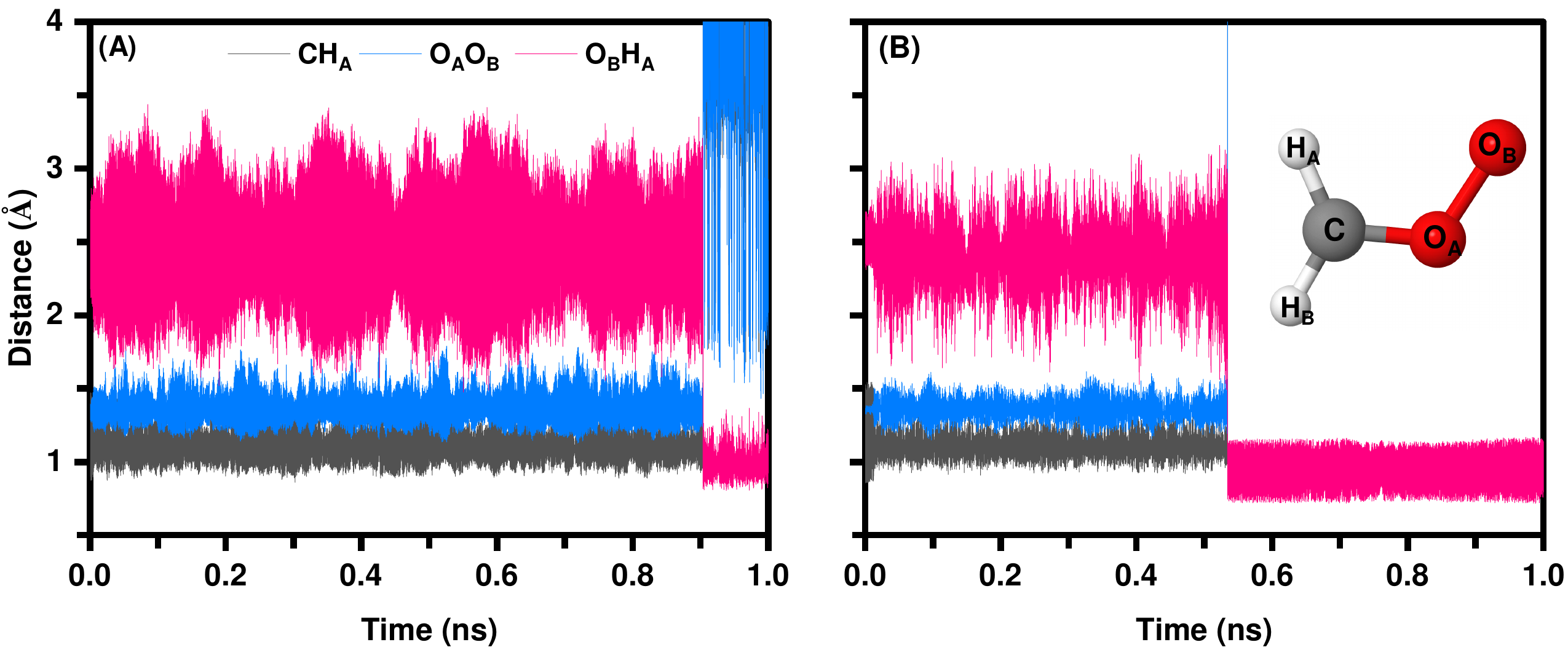}
    \caption{Time series for C\HA{} (black), \OA\OB{} (blue), and
      \OB\HA{} (pink) separations for a reactive trajectory along the
      H-transfer channel with $\sim 4\nu_{\rm CH}$ (32.0 kcal/mol).
      Panel A: Using the PhysNet TL PES H-transfer to form linear
      HCOOH occurs at $t \approx 0.9$ ns (see also Figure
      \ref{fig:timeseries-angle}). Panel B: Using the MS-ARMD PES
      HCOOH forms at $t \approx 0.5$ ns. The amplitudes for the
      \OA\OB{} and \OB\HA{} separations are comparable whereas for the
      C\HA{} bond the PhysNet PES is softer.}
    \label{fig:timeseries-dis}
\end{figure}

\section{Results}
First, the representations of the PESs are validated with respect to
the reference data, followed by a characterization of the reaction
dynamics and the rates for formation of HCOOH and dioxirane.\\

\subsection{Validation of the PESs}
{\bf PhysNet Base Model:} The performance metrics of the base model on
energies and forces for the test set are summarized in Figure
\ref{fig:valid-base}. The MAE$_{\rm train}(E)$ and MAE$_{\rm test}(E)$
are 0.007, 0.009 kcal/mol, and the corresponding RMSE$_{\rm train}(E)$
and RMSE$_{\rm test}(E)$ are 0.019, 0.062 kcal/mol. For the forces the
MAE$_{\rm train}(F)$ and MAE$_{\rm test}(F)$ are 0.022, 0.063
kcal/(mol$\cdot$\AA), and the corresponding RMSE$_{\rm train}(F)$ and
RMSE$_{\rm test}(F)$ are 0.251, 0.597 kcal/(mol$\cdot$\AA),
respectively. Optimized structures for the H$_2$COO, dioxirane, HCOOH
structures and the TSs connecting them agree to within better than an
RMSD of $10^{-3}$ \AA\/ with optimized structures from CCSD(T)-F12
calculations and their energies agree to within better than $10^{-2}$
kcal/mol, see Table \ref{tab:base}. For the harmonic frequencies, the
absolute deviations between the predictions for the 5 stationary
points and the corresponding {\it ab initio} values are all smaller
than 6 cm$^{-1}$, see Figure \ref{fig:freq-ccsd}.\\

\noindent
For following the reaction it is also of interest to report the
minimum energy path and compare the performance of the base model with
reference calculations at the CCSD(T)-F12 level, see Figure
\ref{fig:mep-ccsd-28307}. The RMSD between the reference and PhysNet
energies for the test set is 0.063 kcal/mol. Finally, Diffusion Monte
Carlo (DMC) simulations were carried out to probe the PES for
holes. Propagating 30000 walkers for 50000 steps did not detect a
single hole which illustrates the robustness of the PES.\\

\begin{figure}
    \centering
    \includegraphics[width=\textwidth]{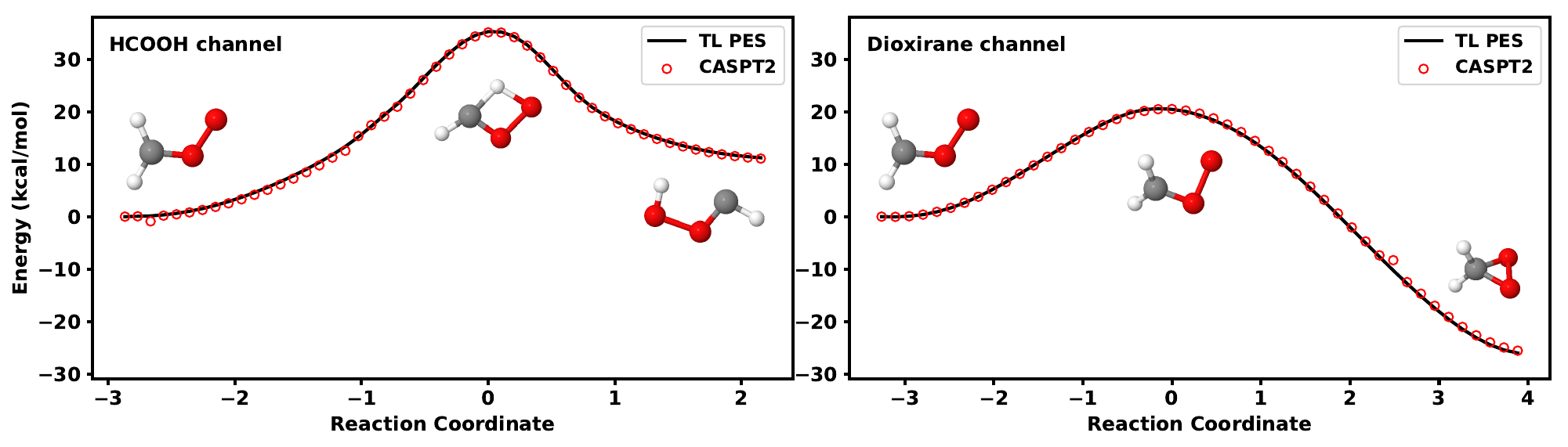}
    \caption{Energy profiles of the H-transfer channel (left panel)
      and the dioxirane formation channel (right panel). Here, the
      black solid line represents the energies from the PhysNet TL
      model, and the red open circles are the reference (CASPT2)
      energies.}
    \label{fig:mep-trained}
\end{figure}

\noindent
{\bf Transfer Learned Model to CASPT2:} The quality of the PhysNet TL
PES is reported in Figure \ref{fig:quality}. The RMSE$(E)$ and
MAE$(E)$ between reference calculations and inference of the NN are
0.85 kcal/mol and 0.41 kcal/mol, respectively. Figure
\ref{fig:valid-tf-5162} provides additional performance metrics. On
energies for the TL model, the MAE$_{\rm train}(E)$ and MAE$_{\rm
  test}(E)$ are 0.43, 0.41 kcal/mol, and the corresponding RMSE$_{\rm
  train}(E)$ and RMSE$_{\rm test}(E)$ are 0.91, 0.85 kcal/mol. The
MAE$_{\rm train}(F)$ and MAE$_{\rm test}(F)$ on forces for the TL
model are 0.39, 0.75 kcal/(mol$\cdot$\AA), and the corresponding
RMSE$_{\rm train}(F)$ and RMSE$_{\rm test}(F)$ are 1.46, 3.54
kcal/(mol$\cdot$\AA). Optimized structures at the CASPT2 level and
from using the TL-PES agree to within a RMSD of 0.01 \AA\/. Finally,
the barrier heights (21.4 and 35.5 kcal/mol for the dioxirane and
HCOOH pathways) from CASPT2 calculations and from using the TL-PES
differ by 0.81 and 0.22 kcal/mol and the minimum energy paths are
reported in Figure \ref{fig:mep-trained}.\\

\begin{figure}
    \centering
    \includegraphics[width=0.8\textwidth]{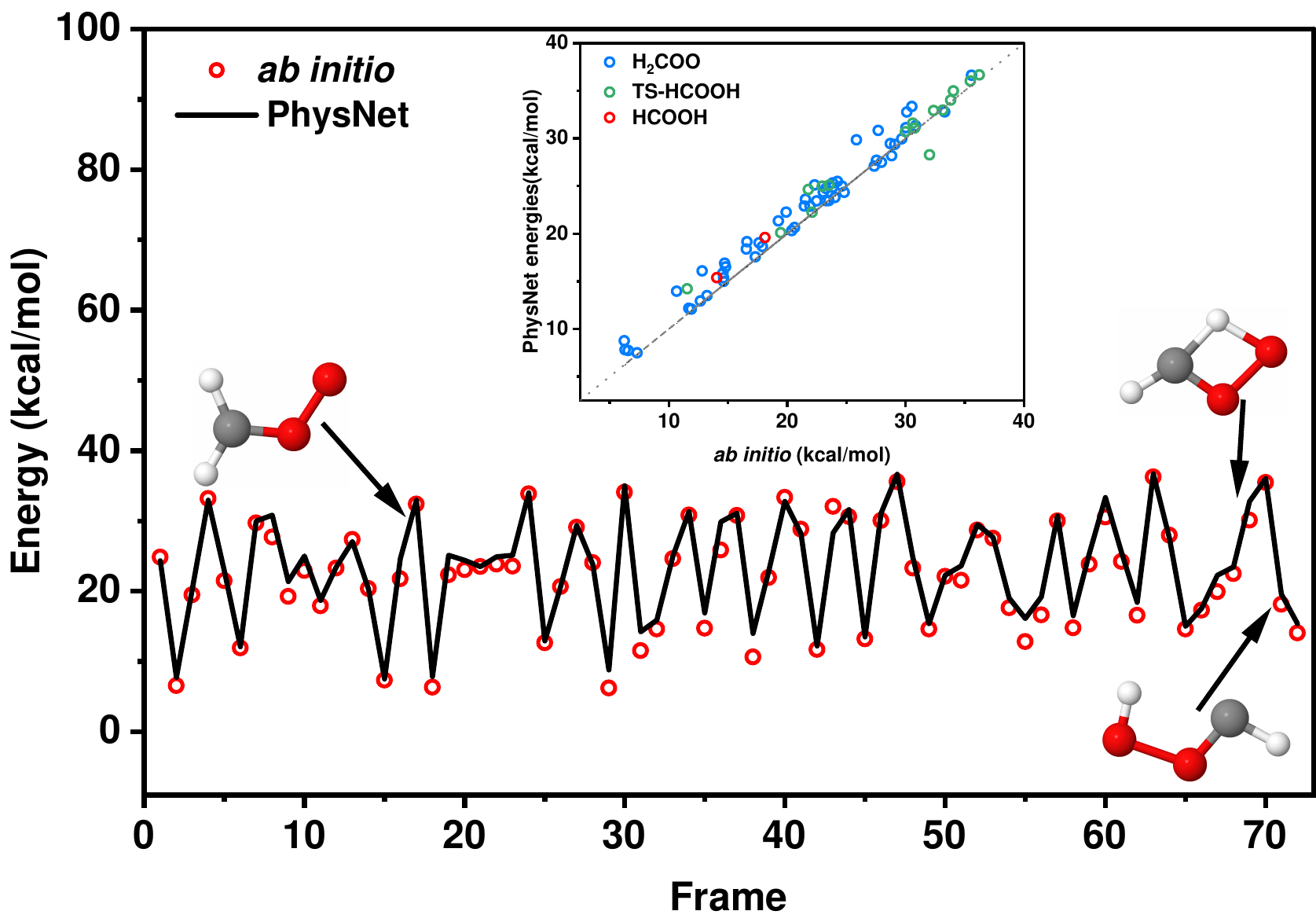}
    \caption{Main view: Comparison of reference CASPT2 energies (red
      circles) and predictions from the PhysNet TL model (black line)
      for a reactive H-transfer trajectory with an excitation energy
      of 32.0 kcal/mol from reactant to product. The inset shows the
      corresponding correlation of reference CASPT2 energies and
      predictions. $R^2$ and RMSE are 0.96 and 1.58 kcal/mol,
      respectively.}
    \label{fig:traj-test}
\end{figure}

\noindent
For a more comprehensive evaluation of the TL model's performance, a
limited set of simulations involving the excitation of the CH stretch
mode with 32.0 kcal/mol was run. For a single reactive trajectory for
the H-transfer channel structures were extracted at regular intervals
between reactant and product. Energies for these structures from
CASPT/aVTZ calculations (red circles) and from the TL-PES (black line)
are compared in Figure \ref{fig:traj-test}. With respect to the
trained NN-PES all these geometries are off-grid and the agreement
between reference and model energies is very encouraging with $R^2 =
0.96$ and ${\rm RMSE} = 1.58$ kcal/mol, respectively.\\

\noindent
{\bf MS-ARMD:} The quality of the MS-ARMD representations of the
reactive PESs are reported in Figure \ref{fig:quality}. The fitted
PESs have root mean squared errors of $\approx 1.0$ kcal/mol and the
IRC closely follows the reference calculations (see
insets). Preliminary simulations were run and energy conservation was
observed for both H-transfer and dioxirane channels. Contrary to the
machine-learned PESs which allows to simultaneously follow both
reaction pathways, MS-ARMD representations can only be used for either
H-transfer or dioxirane formation. Hence, no branching ratios can be
determined from such simulations.\\

\subsection{Reaction Dynamics}
\begin{figure}
    \centering
    \includegraphics[width=\textwidth]{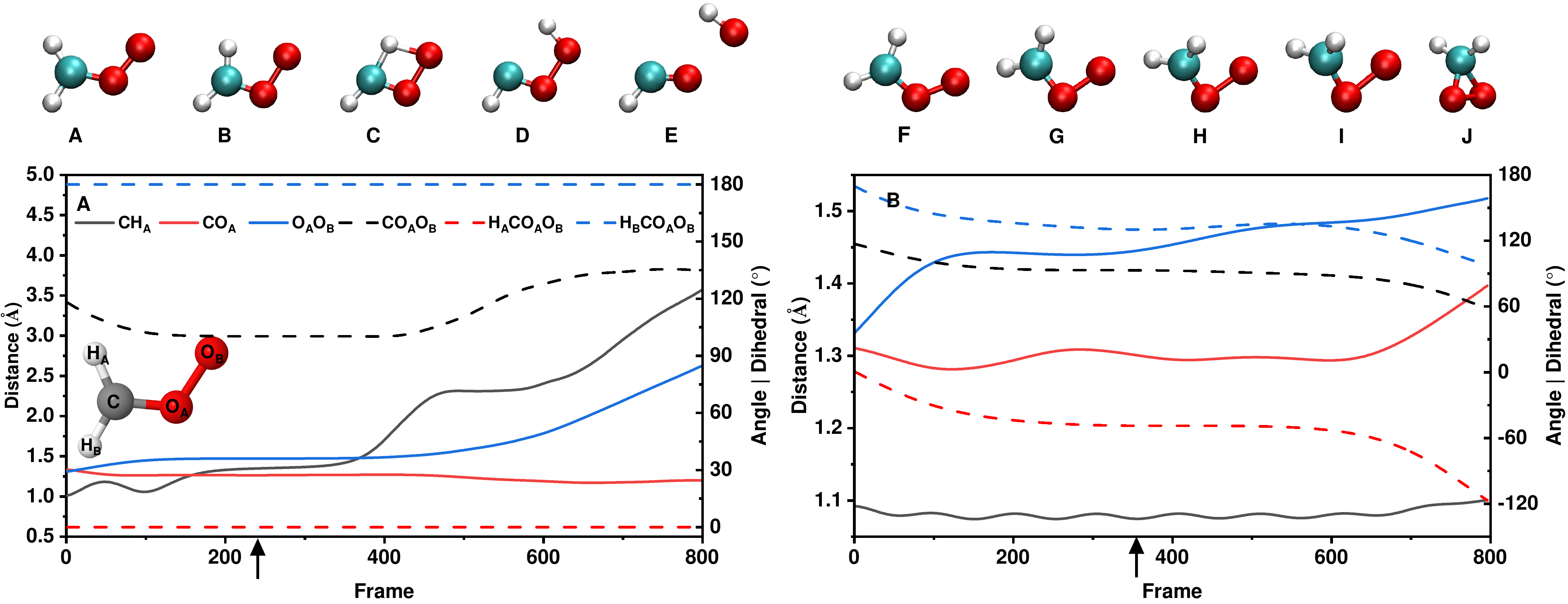}
    \caption{C\HA{}, C\OA{} and \OA\OB{} separations, and change of
      C\OA\OB{} angle, and \HA{}C\OA\OB{} and \HB{}C\OA\OB{} dihedrals
      along the minimum dynamic path from the PhysNet TL PES. Panel A:
      the H-transfer channel from H$_2$COO to HCOOH and then
      dissociation to OH + HCO. Panel B: dioxirane
      formation. Structures A to E and F to J illustrate geometrical
      arrangements along the pathways. The black arrows indicate the
      location of the transition states for the H-transfer (frame 234,
      structure C) and dioxirane (frame 353, structure H) route.}
    \label{fig:mdp}
\end{figure}

For a first impression of the reaction dynamics the minimum dynamic
path (MDP) was determined for the two pathways and using the PhysNet
TL-PES. Such simulations start from the transition state separating
two neighboring minima.\cite{MM.mdp:2019} The excess energies were
$\Delta E = 0.03$ and 0.11 kcal/mol, respectively, and variations of
important internal coordinates along the downhill pathway towards
reactant and product for the ``H-transfer'' and the ``dioxirane''
pathways are reported in Figure \ref{fig:mdp} for the TL-PES.\\

\noindent
For H-transfer the most important participating internal degree of
freedom for the reactant$\rightarrow$TS step as judged from the MDP is
the C\HA{} stretch (solid black). The C\OA{} and \OA\OB{} separations
(solid red and blue) do only change insignificantly in approaching the
TS. Similarly, the two dihedral angles (dashed red and blue) do not
vary and the chemical transformation occurs in a planar
arrangement. There is, however, a readjustment of the C\OA\OB{} angle
(dashed black) from $118^\circ$ to $100^\circ$. These geometrical
changes imply that vibrational excitation along the CH-stretch normal
mode will be most effective to promote reactivity of H$_2$COO towards
the TS leading to the linear HCOOH product and beyond. On the product
side it is interesting to note that the \OA\OB{} distance starts to
increase after passing the TS which facilitates breakup towards
OH-elimination.\\

\noindent
Contrary to that, the dioxirane channel is characterized by an
insignificant change along the C\HA{} bond, some variation of the
C\OA{} separation, and an increase of the \OA\OB{} bond length by 0.1
\AA\/ when moving towards the TS. At the same time, all three angles
considered decrease in concert. Specifically, the C\OA\OB{} angle
changes from $117^\circ$ to $92^\circ$. Taken together, this suggests
that sole excitation along the C\HA{} stretch mode is not expected to
be particularly effective for dioxirane formation but a combination
band involving the C\OA\OB{} bend may be a good and productive
reaction coordinate.\\

\noindent
Consequently, two types of vibrational excitations were
considered. One used exclusively the C\HA{} normal mode for which
excitation energies of 16 kcal/mol to 40 kcal/mol along this vibration
were investigated ($\sim 2 \nu_{\rm CH}$ to $\sim 5 \nu_{\rm CH}$
quanta). This is akin to previous experimental\cite{lester:2016} and
computational\cite{MM.criegee:2021,MM.criegee:2023} work which also
did not employ resonant excitation and the precise number of quanta in
a particular degree of freedom is not essential. The second scheme
used excitation energies of 22.0 kcal/mol and 25.5 kcal/mol along two
CH-stretch/COO-bend combinations: $\sim (3\nu_{\rm CH} + \nu_{\rm
  COO})$ and $\sim (2\nu_{\rm CH} + 4 \nu_{\rm COO})$.\\

\begin{figure}
    \centering
    \includegraphics[width=\textwidth]{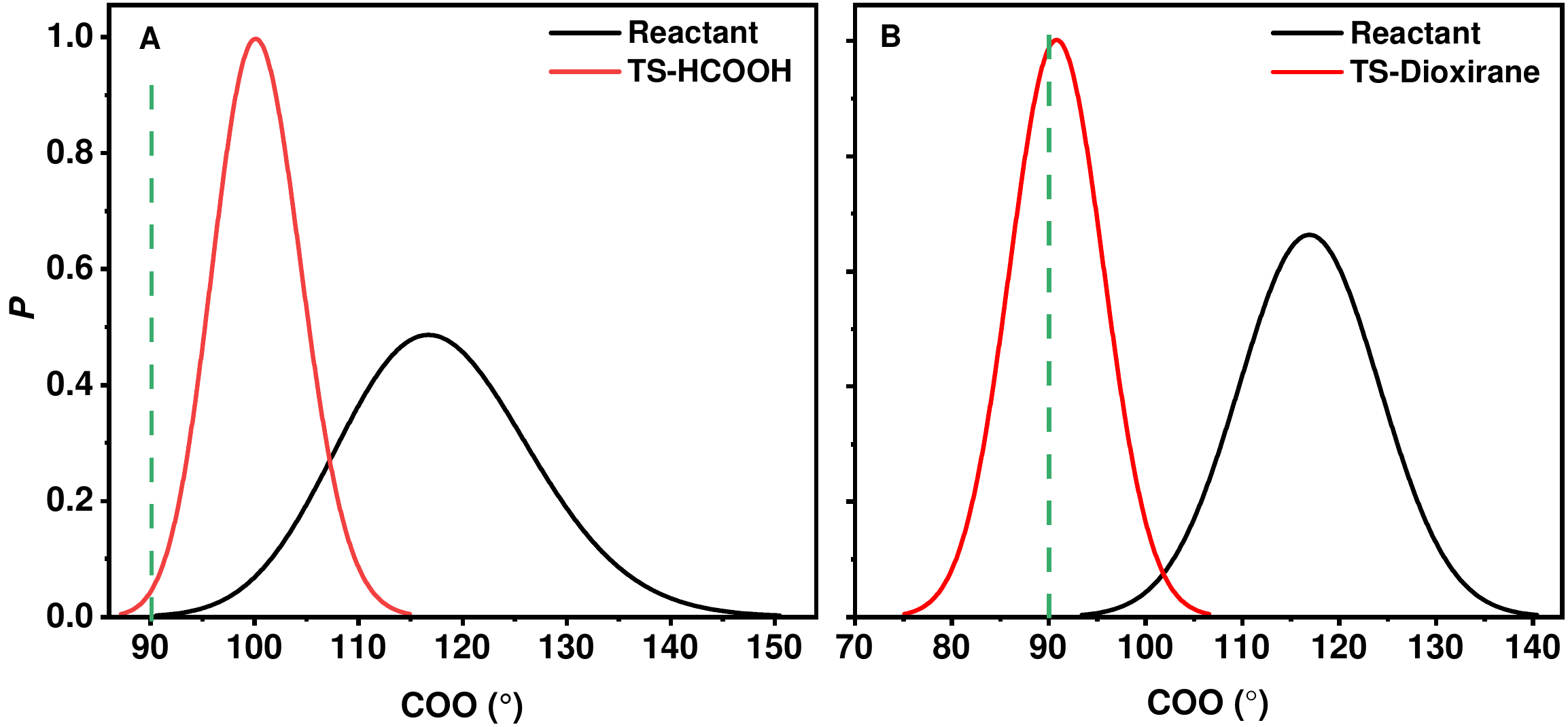}
    \caption{Panel \textbf{A}: Distribution of C\OA\OB{} angle for
      2000 reactive trajectories from the simulations with $\sim
      5\nu_{\rm CH}$ (40.0 kcal/mol) on the PhysNet TL PES for the
      reactant (black) and the transition state (red) for the linear
      pathway. The dashed green line indicates the C\OA\OB{} angle at
      the TS for accessing the dioxirane pathway. As trajectories with
      excitation exclusively along the C\HA{} bond do not sample
      values $\leq 90^\circ$ for the C\OA\OB{} angle only the linear
      pathway leading to HCOOH is followed; see text for
      discussion. Panel \textbf{B}: Distribution of C\OA\OB{} angle
      for 1400 reactive trajectories from the simulations with
      excitation $\sim 3\nu_{\rm CH} + \nu_{\rm COO}$ (25.5 kcal/mol)
      on the PhysNet TL PES for the reactant (black) and the
      transition state (red) for the dioxirane channel. The dashed
      green line indicates the C\OA\OB{} angle at the TS. }
    \label{fig:coo-dist}
\end{figure}

\noindent
Figure \ref{fig:coo-dist}A reports the distribution of C\OA\OB{}
angles following excitation with 40.0 kcal/mol $(\sim 5\nu_{\rm CH})$
and the ensuing dynamics. Excitation along the CH-normal mode yields
almost entirely the linear HCOOH isomer. The distribution functions
for the reactant (black) and the TS towards linear HCOOH (red) clarify
that the geometry of the TS towards dioxirane (green dashed line) is
virtually never sampled. The maximum of the probability distribution
function for the reactant is centered at $120^\circ$ and extends from
$\sim 90^\circ$ to $\sim 150^\circ$. For the transition state towards
linear HCOOH the maximum shifts to $100^\circ$ and the distribution
narrows considerably, extending only between $\sim 90^\circ$ and $\sim
110^\circ$. The positions of the two maxima are also consistent with
the MDP, see Figure \ref{fig:mdp}A. Analysis of these trajectories,
({\it vide infra}) indicates that all of them develop towards the TS
leading to the linear isomer and dioxirane formation is unlikely
despite the fact that the excitation energy of $\sim 40$ kcal/mol is
considerably larger than the barrier towards dioxirane (21.4 kcal/mol,
see Figure \ref{fig:pathway}). One important reason for this is that
the TS leading to dioxirane is characterized by a C\OA\OB{} angle of
$\sim 90^\circ$ (green dashed line in Figure \ref{fig:coo-dist}) which
is unlikely to be sampled for trajectories in which vibrational
excitation occurs along the CH-stretch normal mode only. In other
words, using the TL-PES energy transfer between the CH-stretching and
the COO-bending motion is ineffective on the time scale of the present
simulations (1 ns).\\

\begin{figure}
    \centering
    \includegraphics[width=0.8\textwidth]{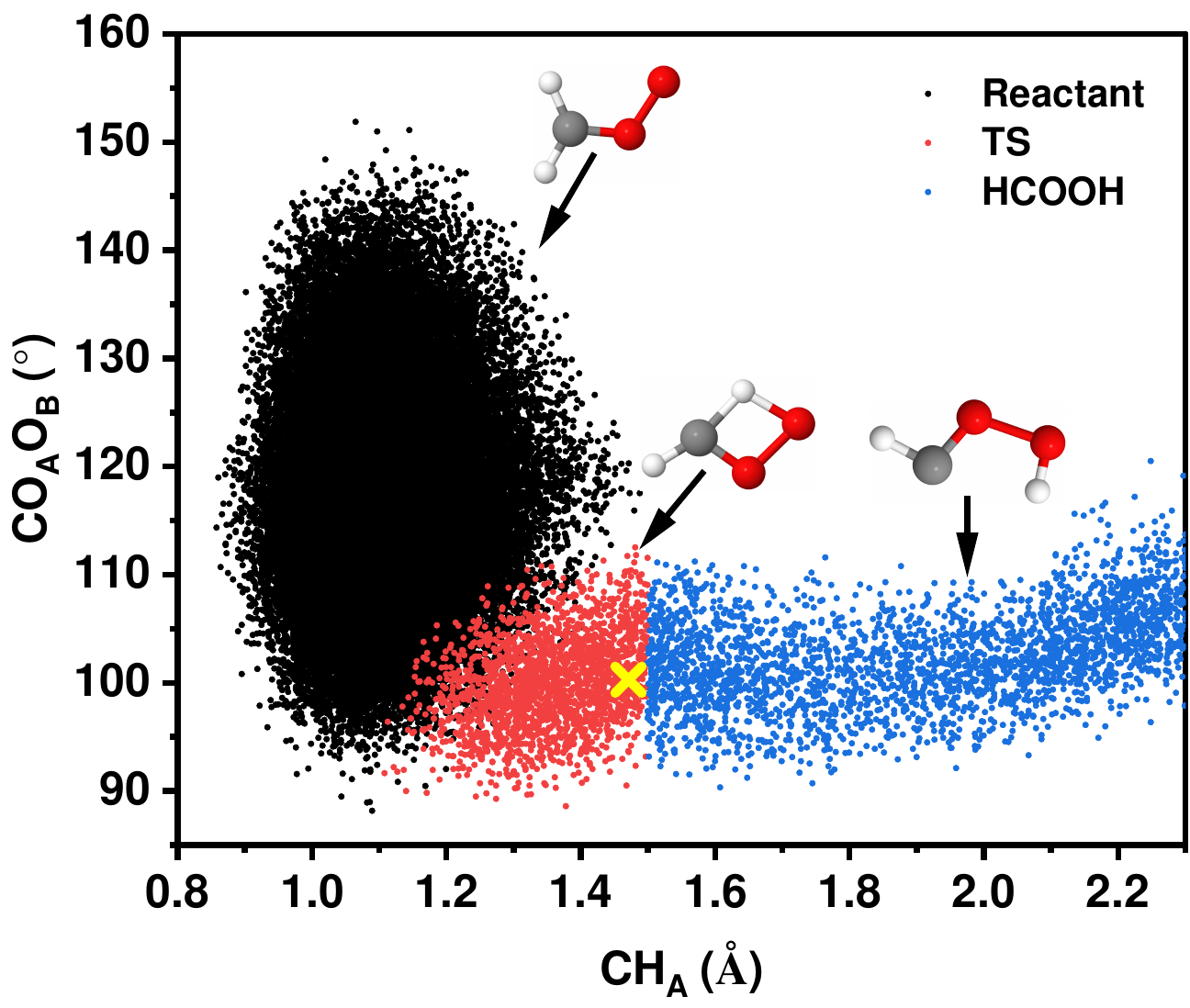}
    \caption{Correlation between the C\OA\OB{} angle and the C\HA{}
      separation for 2500 reactive trajectories from the simulations
      with 40.0 kcal/mol $(\sim 5 \nu_{\rm CH})$ for the H-transfer
      channel on the PhysNet TL PES. The width of the transition seam
      along the \OA\OB{}-separation is 0.4 \AA. The yellow cross is at
      the TS geometry from the CASPT2/aVTZ calculations.}
    \label{fig:2dhcooh}
\end{figure}

\noindent
Because the MDP for the linear pathway (see Figure \ref{fig:mdp}A)
indicated that the C\HA-stretch changes appreciably and the
C\OA\OB-bend also varies in approaching the TS, it is of interest to
characterize their correlated motion along the reaction pathway. For
this, the geometries sampled in the reactant, TS, and product
geometries were separately analyzed following excitation with 40.0
kcal/mol, see Figure \ref{fig:2dhcooh}. The black (reactant), red
(TS), and blue (HCOOH) point clouds provide a comprehensive
description of the motions that are followed for the H-transfer
reaction path and clarify that the transition region has finite width,
both in the C\HA-stretch and the C\OA\OB-bend coordinates, extending
well beyond the transition state (yellow cross). \\

\begin{figure}
    \centering
    \includegraphics[width=\textwidth]{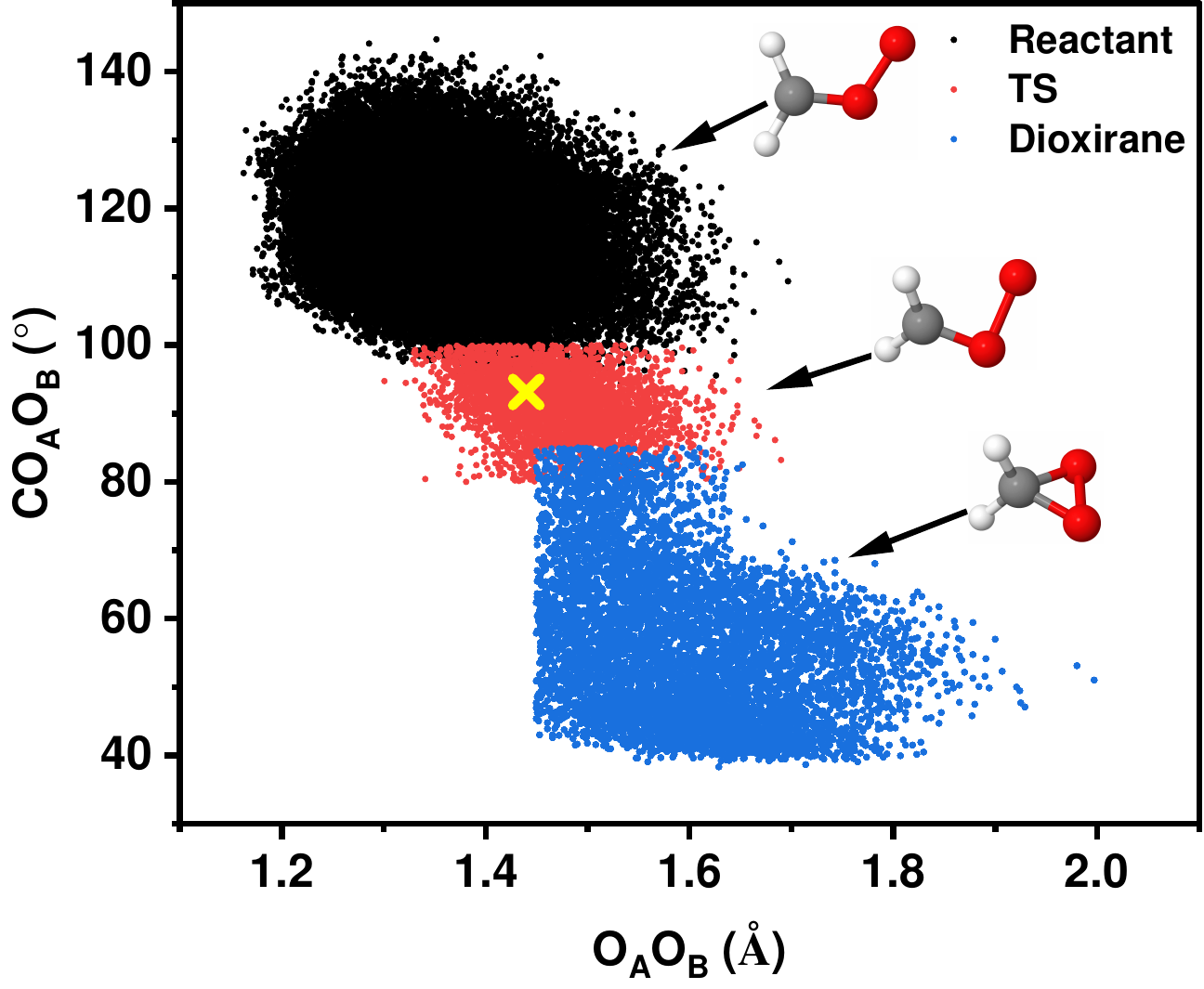}
    \caption{Distribution of C\OA\OB{} angle as a function of the
      \OA\OB{} distance for 1400 reactive trajectories from
      simulations with 25.5 kcal/mol ($\sim 3\nu_{\rm CH} + \nu_{\rm
        COO}$) for the dioxirane channel on the PhysNet TL PES. The
      width of the transition seam along the \OA\OB{}-separation is
      0.4 \AA. The yellow cross represents the TS geometry from
      CASPT2/aVTZ calculations.}
    \label{fig:distr-dioxirane}
\end{figure}

\noindent
Following excitation of the combination mode with 25.5 kcal/mol,
corresponding to $\sim (3\nu_{\rm CH} + \nu_{\rm COO})$, only the
dioxirane pathway is followed. The C\OA\OB-distribution functions
$P(\theta_{\rm COO})$ for the reactant (black) and the TS-geometries
(red) leading to dioxirane are reported in Figure \ref{fig:coo-dist}B
together with the C\OA\OB{} angle for the TS from the electronic
structure calculations (green dashed line). Evidently, the dynamics
samples the TS-geometry extensively and the sampled widths of the
distributions decrease in going from the reactant to the TS whereby
the TS structure for the linear pathway is only rarely sampled. The
correlation between \OA\OB-stretch and the C\OA\OB-bend coordinates
for this pathway is shown in Figure \ref{fig:distr-dioxirane}. It is
interesting to note that product state geometries for the \OA\OB-bond
lengths are already sampled in the reactant well and the decisive
coordinate along the dioxirane-pathway is the C\OA\OB-bend.\\

\subsection{Reaction Probabilities}
As mentioned earlier, the excitation schemes considered in the present
work are chosen to follow the H-transfer or dioxirane-formation
pathways. Excitation along the CH-stretch normal mode yields
predominantly linear HCOOH and only minor amounts of dioxirane even
for the highest excitation energy of 40.0 kcal/mol despite a barrier
height of only 21.4 kcal/mol to form dioxirane. Conversely, excitation
of the CH-stretch/COO-bend combination mode leads exclusively to
dioxirane primarily because energies of 22.0 to 25.5 kcal/mol are not
sufficient to reach the TS energy towards HCOOH of 33.9 kcal/mol.\\

\begin{table}[h!]
\centering
\caption{Reaction probability for H-transfer channel from simulations
  by exciting the CH stretch mode with PhysNet TL PES and MS-ARMD
  force field.}
\begin{tabular}{|l|c|c|c|c|}
\hline
\begin{tabular}[c]{@{}l@{}}Excitation Energy\\ (kcal/mol)\end{tabular} & \begin{tabular}[c]{@{}c@{}}$\sim 2$ quanta\\ (16.0)\end{tabular} & \begin{tabular}[c]{@{}c@{}}$\sim 3$ quanta\\ (24.0)\end{tabular} & \begin{tabular}[c]{@{}c@{}}$\sim 4$ quanta\\ (32.0)\end{tabular} & \begin{tabular}[c]{@{}c@{}}$\sim 5$ quanta\\ (40.0)\end{tabular} \\ \hline
PhysNet ($E_{\rm TS}^{\rm CASPT2} = 35.5$ kcal/mol)      & 0        & 0        & 0.7 \%      & 98.4 \%        \\ \hline
MS-ARMD ($E_{\rm TS}^{\rm CCSD(T)} = 33.9$ kcal/mol)   & 0        & 0        & 3.0 \%      & 96.0 \%       \\ \hline
\end{tabular}
\label{tab:hcooh}
\end{table}

\begin{figure}
    \centering
    \includegraphics[width=\textwidth]{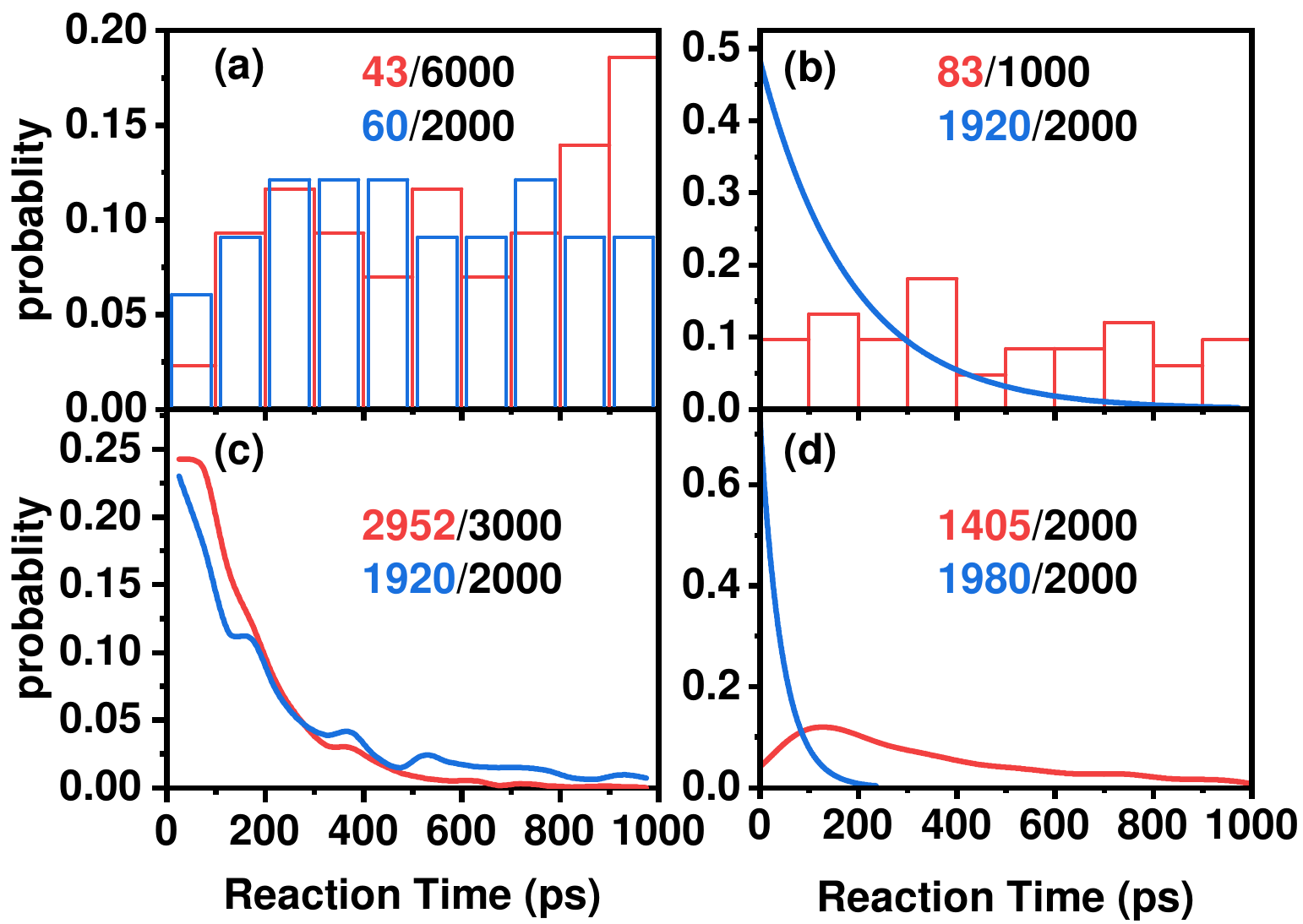}
    \caption{Normalized reaction time distribution from the MD
      simulations for the H-transfer (left column) and dioxirane
      (right column) pathways using the TL-PES (red) and MS-ARMD
      (blue). Excitation energies in panels a to d are 32.0 kcal/mol
      ($\sim 4\nu_{\rm CH }$), 22.0 kcal/mol ($\sim (2\nu_{\rm CH}$ +
      $4\nu_{\rm COO})$), 40.0 kcal/mol ($\sim 5\nu_{\rm CH }$), and
      25.5 kcal/mol ($\sim (3\nu_{\rm CH }$ + $1\nu_{\rm COO}$)),
      respectively.}
     \label{fig:time_dis}
\end{figure}

\noindent
Table \ref{tab:hcooh} shows that for excitation energies below or
close to the barrier height for the H-transfer pathway (33.9 kcal/mol
at CCSD(T)/aVTZ and 35.5 kcal/mol at CASPT2/aVTZ) the reaction
probability is 0 or a few percent. Excitation with 40.0 kcal/mol
$(\sim 5 \nu_{\rm CH})$ leads to formation of HCOOH in almost all
cases. Both, the reaction probabilities and the distribution of
reaction times agree well for simulations using MS-ARMD and the
TL-PES. Because excitation is close to the barrier height with 32.0
kcal/mol $(\sim 4 \nu_{\rm CH})$ the reaction times $\tau$ are widely
distributed, see Figure \ref{fig:time_dis}a, and converging them would
require a considerably larger number of trajectories. Contrary to
that, exciting the CH-stretch normal mode with 40.0 kcal/mol shifts
the maximum of the reaction times to short $\tau$ (Figure
\ref{fig:time_dis}c).\\

\noindent
Following the dioxirane-formation channel features a barrier height of
20.0 and 21.4 kcal/mol at the CCSD(T)/aVTZ and CASPT2/aVTZ levels of
theory, respectively, and requires excitation of the
CH-stretch/COO-bend combination band. With an excitation energy of
22.0 kcal/mol $(\sim (2\nu_{\rm CH}$ + $4\nu_{\rm COO}))$ 96.0 \% of
the trajectories yield dioxirane if the MS-ARMD PES is used compared
with 8.3 \% when running the simulations with the TL-PES. Part of this
difference arises due to the lower barrier on the MS-ARMD
PES. Excitation with 25.5 kcal/mol features 99.0 \% compared with 70.3
\% of reactive trajectories, see Table \ref{tab:dioxy}. The reaction
time distributions for the two PESs differ quite substantially. Using
the MS-ARMD PES, $P(\tau)$ for excitation with 22.0 kcal/mol peaks at
50.5 ps (blue trace in Figure \ref{fig:time_dis}d) compared to 159.2
ps with the TL-PES (red trace in Figure
\ref{fig:time_dis}d). Excitation with 25.5 kcal/mol broadens all
distributions and a multimodal structure for $P(\tau)$ appears. Again,
there is a pronounced peak at early reaction times for simulations
using MS-ARMD whereas with TL-PES this feature is missing. Most likely
this pronounced peak at short reaction times is due to the lower
barrier height of the MS-ARMD PES by 1.4 kcal/mol compared with the
TL-PES.\\

\begin{table}[h!]
\centering
\caption{Reaction probability for dioxirane channel from simulations
  by exciting the CH stretch mode and the COO bending mode with
  PhysNet base PES, TL PES and MS-ARMD force field.}
\begin{tabular}{|l|c|c|}
\hline
\begin{tabular}[c]{@{}l@{}}Excitation Energy\\ (kcal/mol)\end{tabular} &
\begin{tabular}[c]{c@{}lc@{}}$\sim (2\nu_{\rm CH} + 4\nu_{\rm COO})$\\ (22.0)\end{tabular} &
\begin{tabular}[c]{c@{}lc@{}}$\sim (3\nu_{\rm CH} + 1\nu_{\rm COO})$\\(25.5)\end{tabular} \\ \hline
PhysNet ($E_{\rm TS} ^{\rm CCSD(T)} = 20.0$ kcal/mol)                                     & 20.3 \%     & 67.1 \%          \\ \hline
PhysNet ($E_{\rm TS}^{\rm CASPT2} = 21.4$ kcal/mol)                                      & 8.3 \%      & 70.3 \%  \\ \hline
MS-ARMD ($E_{\rm TS}^{\rm CCSD(T)} = 20.0$ kcal/mol)                                      & 96.0 \%  & 99.0 \%  \\ \hline
\end{tabular}
\label{tab:dioxy}
\end{table}

\section{Discussion and Conclusion}
The present work reports on the reactive dynamics following
vibrational excitation of the smallest Criegee intermediate,
H$_2$COO. Excitation of vibrations to drive atmospheric chemical
reactions has been use for {\it
  syn-}CH$_3$COOH.\cite{lester:2016,MM.criegee:2021,MM.criegee:2023}
and considered for OH-elimination from species such as HSO$_3$X (X =
F, Cl, OH).\cite{MM.hso3f:2017,Feierabend:2006} Contrary to these
examples, for H$_2$COO two reaction pathways were considered which
lead to different product channels. Importantly, the present work
suggests that depending on the vibrational modes that are excited, the
two pathways are almost selectively followed from simulations using
the TL-PES which allows to access both pathways in the same
simulation. If the CH-stretch normal mode is excited with $\sim 5
\nu_{\rm CH}$ (40.0 kcal/mol), the probability for H-transfer and
formation of linear HCOOH is two orders of magnitude larger than
dioxirane formation. This is despite the fact that the barrier height
for formation of dioxirane is only 21.4 kcal/mol at the CASPT/aVTZ
level of theory. Conversely, excitation of the COO-bend with energy
equivalent to 1 quantum in the $\nu_{\rm COO}$ mode is sufficient to
yield appreciable amounts of dioxirane using both PESs considered. In
other words, mode selective chemistry is observed for H$_2$COO.\\

\noindent
The possibility to deposit energy corresponding to multiple quanta in
stretch modes was demonstrated for H$_2$SO$_4$ for which the
4${\nu}_9$ and 5${\nu}_9$ O-H stretching overtones were
excited.\cite{Feierabend:2006} Similarly, vibrational ``ladder
climbing'' was used to deposit up to 4 quanta of vibrational energy in
nitrile (-CN) functionalized phenol.\cite{kraack:2016} Finally,
excitation of high vibrational states $(v \ge 7)$ of CO-ligands in
Cr(CO)$_6$ was found to lead to CO-dissociation from the parent
molecule.\cite{witte:2003} These findings illustrate that excitation
of highly excited vibrational states is possible to probe the
spectroscopy and even induce reactivity.\\

\noindent
Reactive MD simulations were also run using the NN-base model
determined from CCSD(T)-F12a/aVTZ reference calculations. Although the
quality of the NN-representation is excellent (see e.g. Table
\ref{tab:base}), excitation along the CH-stretch normal mode often
feature problems: for example, C-O bond breaking to form CH$_2$+O$_2$
is found which should not occur at these excitation energies. In
search for a reason it was found that the CCSD(T)-F12a/aVTZ reference
data on the singlet PES feature a local minimum for elongations along
the C-O stretch coordinate at 1.75 \AA\/, corresponding to an energy
of $\sim 30$ kcal/mol. Although the NN-base model reliably captures
this feature, breaking of the CO bond along the singlet-PES at such
low energies is not realistic. The reason for this spurious feature is
the proximity of the triplet-PES which complicates the electronic
structure calculations in this region of configuration space and
single-reference methods are not sufficiently reliable.\\

\noindent
Using TL based on CASPT2/aVTZ improves the situation in that the
spurious minimum is shifted to longer CO-separations and to $\sim 85$
kcal/mol which is considerably higher than the transition state towards
H-transfer at 33.8 kcal/mol. This allows to run meaningful simulations
for the processes of interest in the present work. Nevertheless, no
globally valid PES is yet available and the CASPT2/aVTZ level of
theory does not provide reference data to sufficiently high energies
for developing such a PES. For the dioxirane pathway the situation is
somewhat better and meaningful simulations for the base model are
possible.\\

\noindent
From the perspective of atmospheric chemistry, OH-formation from
vibrational excitation of H$_2$COO is most relevant. The present work
suggests that excitation of the CH-stretch with $\sim 4$ to $\sim 5
\nu_{\rm CH}$ is most likely to yield OH after H-transfer to form
linear HCOOH. This is comparable to the requirements for decomposition
of H$_2$SO$_4$ through excitation of the
OH-stretch.\cite{Donaldson:2003,reyes.pccp.2014.msarmd,Miller:2006}
The pathway through dioxirane involves barriers that are comparable to
the stabilization energy of dioxirane itself, hence IVR will limit
productive OH-generation, see Figure \ref{fig:pathway}. Finally, the
pathway through FA which was not considered here, first involves a
very loose transition state (with O-separations of $\sim 4$ \AA\/ from
the H$_2$CO-core at the CASSCF level). Such a TS is very
``vulnerable'' in an atmospheric environment. Secondly, although the
insertion product (c-FA) is highly stabilized by $-120$ kcal/mol,
subsequent barriers are still of the order of 60 to 80
kcal/mol. Thermodynamically, surmounting such barriers is
feasible. However, again, IVR and collisional de-excitation limit this
process as has, e.g., been shown for acetaldehyde to form
vinoxy-radical.\cite{MM.atmos:2020}\\

\noindent
In summary, the present work explores the reactive dynamics following
vibrational excitation of the smallest Criegee intermediate H$_2$COO
at the CCSD(T) and CASPT2 levels of theory. Two different excitation
schemes almost exclusively lead to reactive dynamics along the
H-transfer and dioxirane-formation pathways. The most promising route
for OH-formation involves excitation of the CH-stretch normal mode
with energies equivalent to between 4 and 5 quanta to yield HCOOH with
subsequent breaking of the O--O bond. Although the molecule only
contains 5 atoms its electronic structure is challenging and despite
the considerable effort in the present and previous work no globally
valid, reactive PES is available as of now.\\

\section*{Supporting Information Available}
The PhysNet codes are available at
\url{https://github.com/MMunibas/PhysNet}, and the PhysNet PESs and
the data sets containing the reference data can be obtained from
\url{https://github.com/MMunibas/H2COO-PhysNet}.

\section*{Acknowledgment}
We thank Prof. Jun Li for providing the datasets for H$_2$COO and
dioxirane. Valuable discussions with Silvan K\"aser are also
acknowledged. This work was supported by China Scholarship Council (to
KS), the Swiss National Science Foundation through grants
$200020\_219779$ and $200021\_215088$ and the University of Basel.

\bibliography{refs.clean}

\clearpage

\renewcommand{\thetable}{S\arabic{table}}
\renewcommand{\thefigure}{S\arabic{figure}}
\renewcommand{\thesection}{S\arabic{section}}
\renewcommand{\d}{\text{d}}
\setcounter{figure}{0}  
\setcounter{section}{0}  
\setcounter{table}{0}

\newpage

\noindent
{\bf SUPPORTING INFORMATION: OH-Formation Following Vibrationally
  Induced Reaction Dynamics of H$_2$COO}

\begin{figure}[h!]
    \centering
    \includegraphics[width=0.8\textwidth]{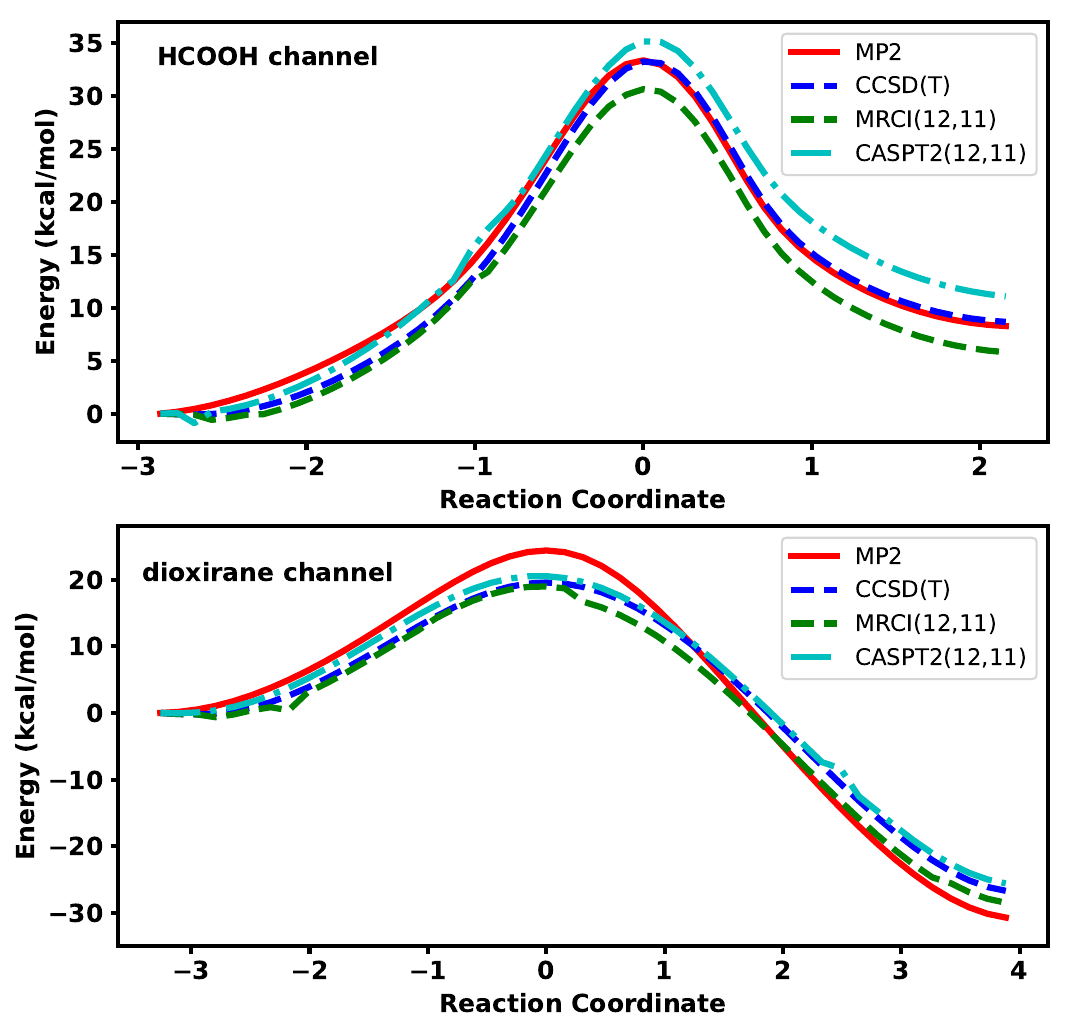}
    \caption{The energy profiles of H$_2$COO for the H-transfer
      channel (upper panel) and for the dioxirane channel (bottom
      panel). The performance of three different level of theory (MP2,
      CCSD(T), MRCI, CASPT2) are included for comparison.}
    \label{fig:mep}
\end{figure}

\begin{figure}
    \centering \includegraphics[scale=0.5]{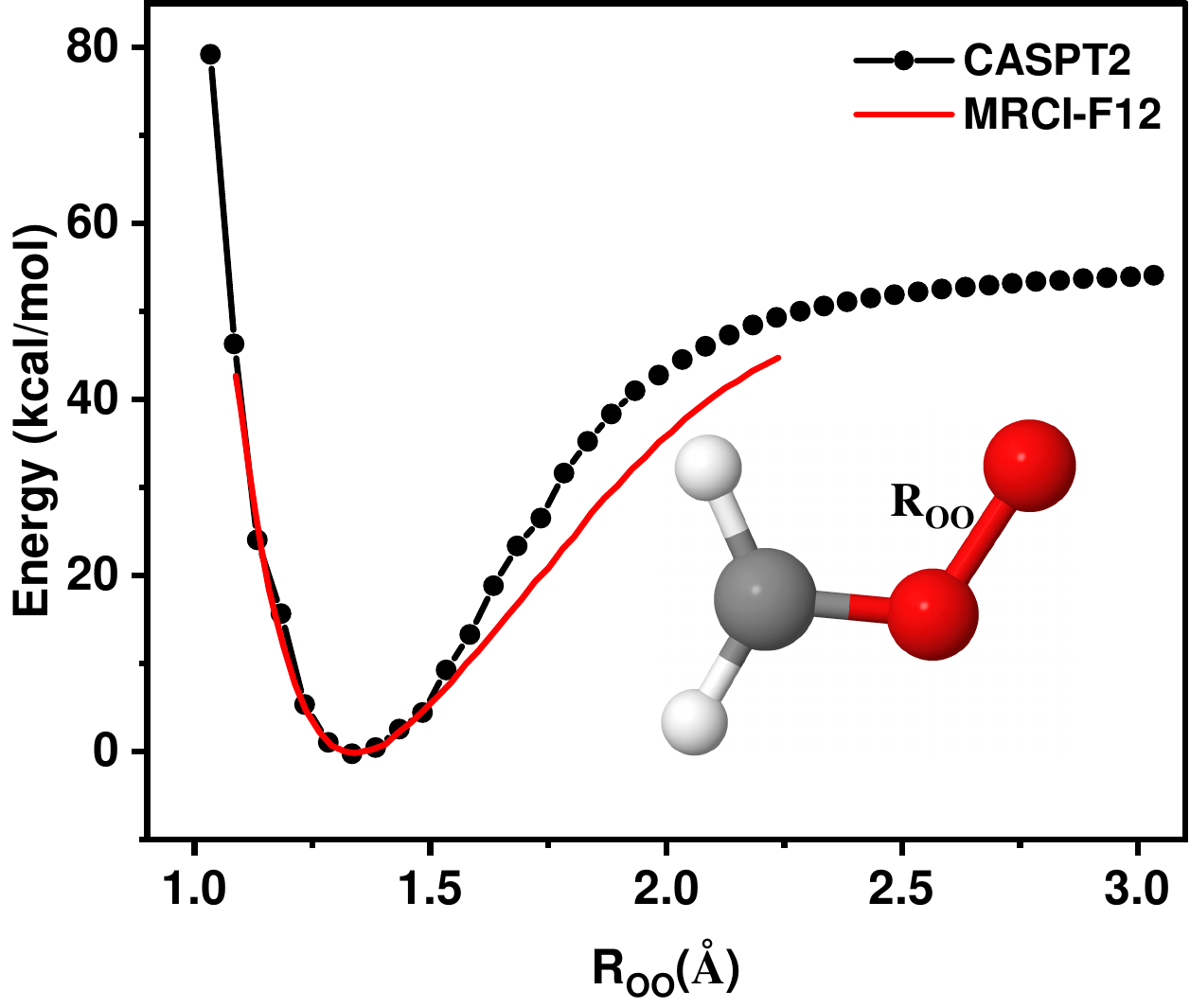}
    \caption{One-dimensional cut at CASPT2 level of theory for the
      reactant along the O-O bond with other coordinates fixed at the
      equilibrium of H$_2$COO. The relaxed scan along the O-O bond
      (red line) at the MRCI-F12 level from Ref.\cite{dawes2015uv} is
      included for comparison. For this, the data was extracted from
      Figure 1 of Ref. \cite{dawes2015uv} using
      WebPlotDigitizer\cite{Rohatgi2022} for graphical
      representation.}
    \label{fig:scan-oo-ch}
\end{figure}

\begin{figure}
    \centering
    \includegraphics[width=0.5\textwidth]{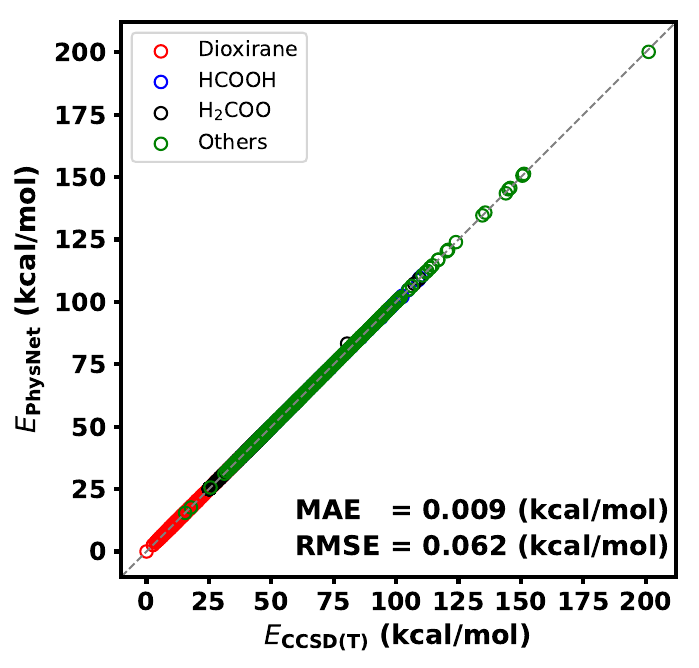}
    \caption{Correlation of 2962 (10\%) \textit{ab initio} energies
      and predicted energies on the test set from the PhysNet base
      model.}
    \label{fig:corr-ccsdt}
\end{figure}  

\begin{figure}
    \centering
    \includegraphics[width=0.8\textwidth]{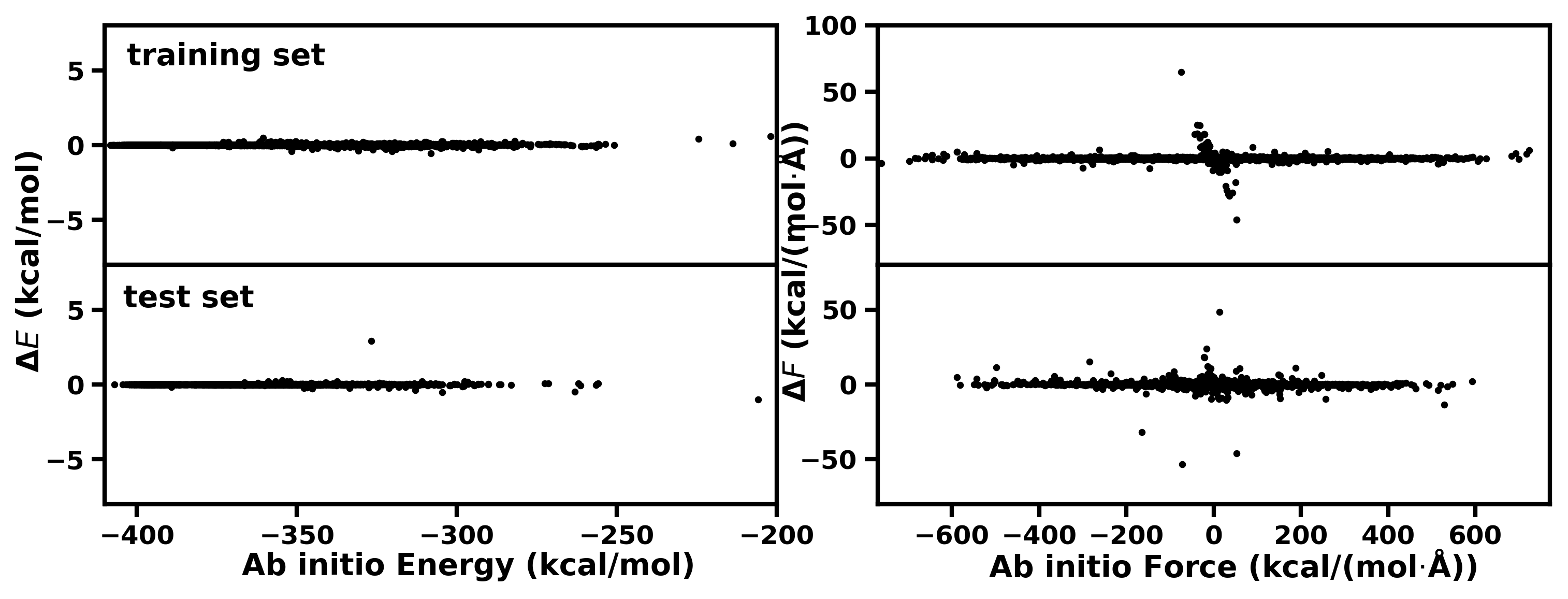}
    \caption{Comparison between reference CCSD(T)-F12a/aVTZ
      energies/forces and predicted energies/forces on the training
      and test sets from the PhysNet base model, from top to
      bottom. The performance of the best PhysNet base model for
      H$_2$OOC is shown. Here, $\Delta E = E_{\rm PhysNet} - E_{\rm
        CCSD}$, $\Delta F = F^{\alpha}_{\rm PhysNet} - F^{\alpha}_{\rm
        CCSD}$ where $\alpha = (x,y,z)$ are the three Cartesian
      components of the forces on each atom. On the energies for the
      base model, the MAE$_{\rm train}(E)$ and MAE$_{\rm test}(E)$ are
      0.007, 0.009 kcal/mol, and the corresponding RMSE$_{\rm
        train}(E)$ and RMSE$_{\rm test}(E)$ are 0.019, 0.062
      kcal/mol. The MAE$_{\rm train}(F)$ and MAE$_{\rm test}(F)$ on
      forces for the model are 0.022, 0.063 kcal/(mol$\cdot$\AA), and
      the corresponding RMSE$_{\rm train}(F)$ and RMSE$_{\rm test}(F)$
      are 0.251, 0.597 kcal/(mol$\cdot$\AA). }
    \label{fig:valid-base}
\end{figure}

\begin{figure}
    \centering
    \includegraphics[width=0.6\textwidth]{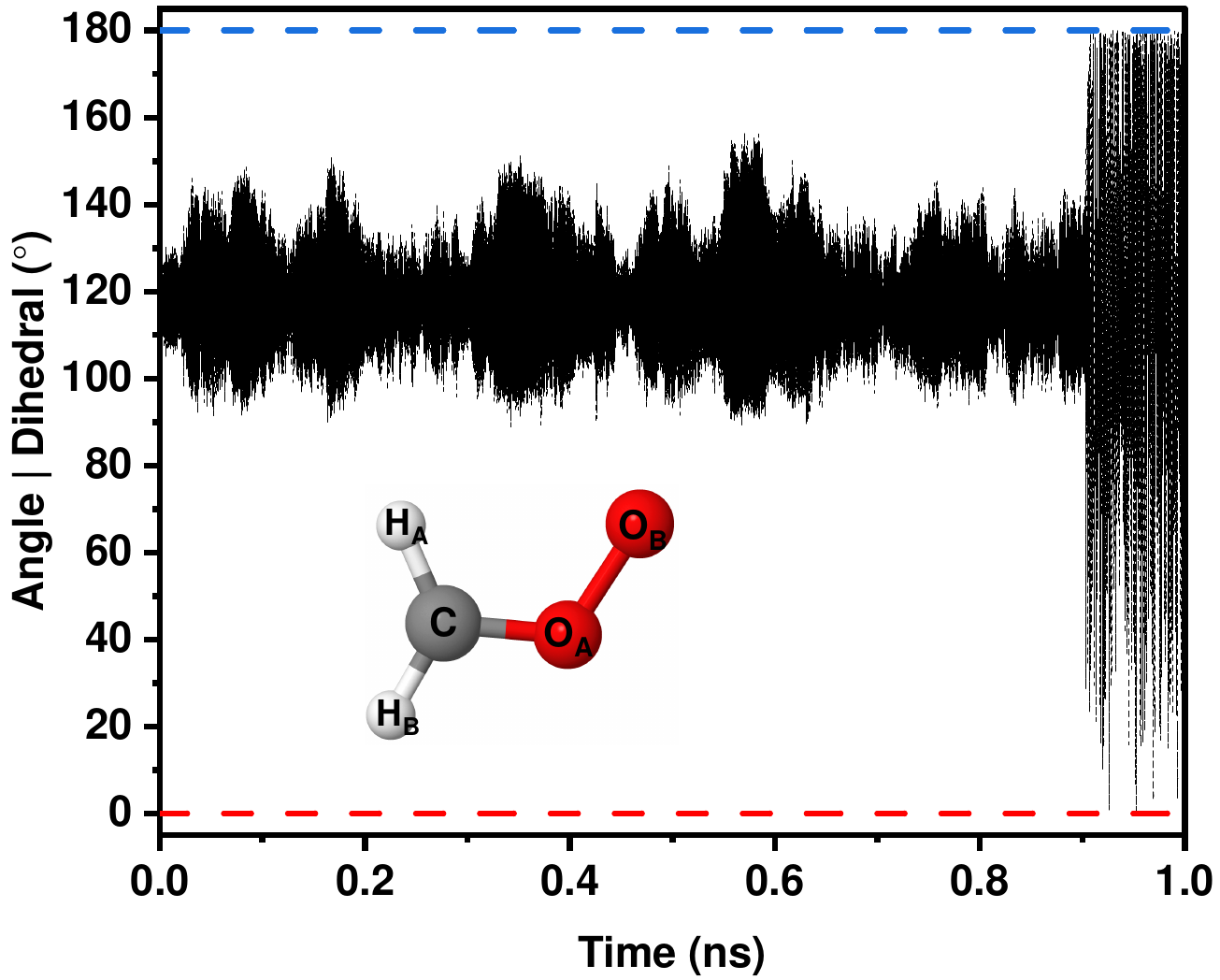}
    \caption{Time series for C\OA\OB{} valence angle (black) and
      \HA{}C\OA\OB{} (red) and \HB{}C\OA\OB{} dihedrals (blue) for a
      reactive trajectory for the H-transfer channel by exciting $\sim
      4 \nu_{\rm CH}$ (32.0 kcal/mol) using the PhysNet TL
      PES. H-transfer occurs after $\sim 0.9$ ns. The molecule remains
      planar throughout the simulation.}
    \label{fig:timeseries-angle}
\end{figure}

\begin{figure}
    \centering
    \includegraphics[width=\textwidth]{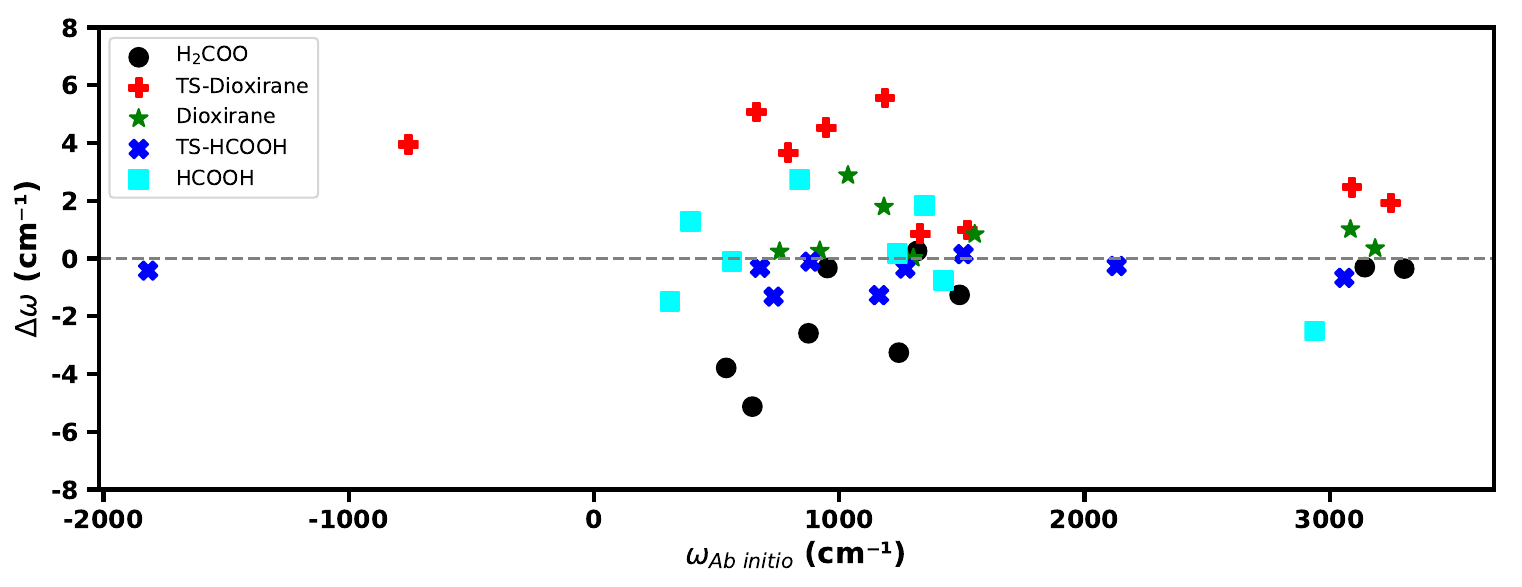}
    \caption{Accuracy of the harmonic frequencies from the PhysNet
      base model is shown, with respect to the appropriate reference
      \textit{ab initio} values. Here, $\Delta\omega = \omega_{\rm
        Ab\ initio} – \omega_{\rm PhysNet}$. All absolute deviations
      of the harmonic frequencies are smaller than 6 cm$^{-1}$. The
      measured experimental (anharmonic) frequencies are 847.44,
      909.26, 1213.30, 1285.90 and 1434.10 cm$^{-1}$, compared with
      872.20, 951.50, 1239.90, 1318.10 and 1490.0 cm$^{-1}$ from the
      base model.\cite{yu2015vibrational,huang2015infrared} }
    \label{fig:freq-ccsd}
\end{figure}

\begin{figure}
    \centering
    \includegraphics[width=\textwidth]{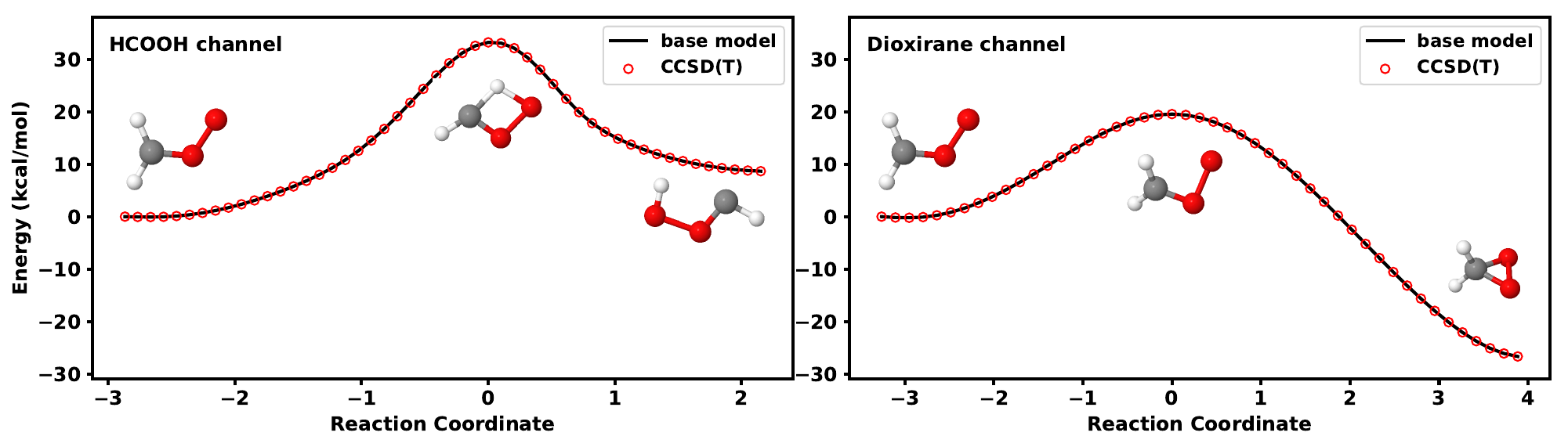}
    \caption{Energy profiles of the H-transfer channel (left panel)
      and the dioxirane formation channel (right panel). Here, the
      black solid line represents the energies from the PhysNet base
      model, and the red open circles refer to the reference
      CCSD(T)-F12a/aVTZ energies. }
    \label{fig:mep-ccsd-28307}
\end{figure}

\begin{figure}
    \centering
    \includegraphics[width=\textwidth]{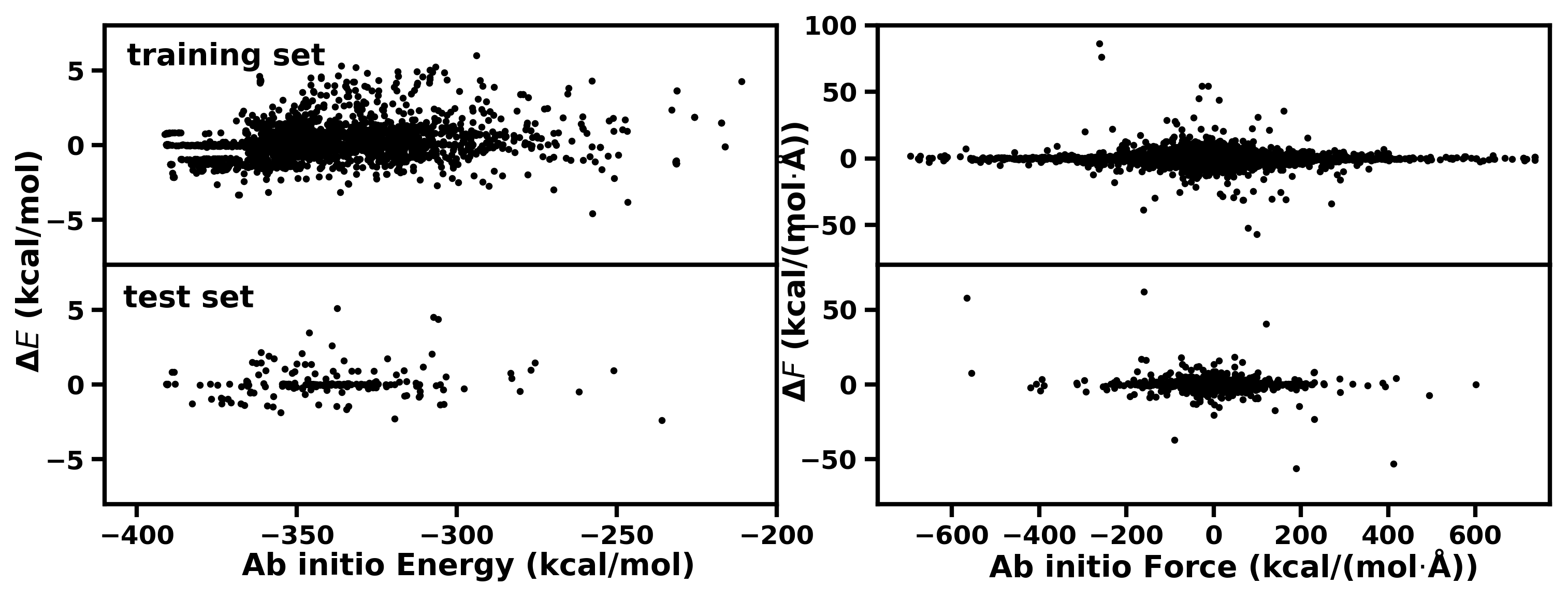}
    \caption{Comparison between reference CASPT2/aVTZ energies/forces
      and predicted energies/forces on the training and test sets from
      the PhysNet TL model, from top to bottom. The performance of the
      best PhysNet TL model for H$_2$OOC is shown. Here, $\Delta E =
      E_{\rm PhysNet} - E_{\rm CCSD}$, $\Delta F = F^{\alpha}_{\rm
        PhysNet} - F^{\alpha}_{\rm CCSD}$ where $\alpha = (x,y,z)$ are
      the three Cartesian components of the forces on each atom. On
      the energies for the TL model, the MAE$_{\rm train}(E)$ and
      MAE$_{\rm test}(E)$ are 0.43, 0.41 kcal/mol, and the
      corresponding RMSE$_{\rm train}(E)$ and RMSE$_{\rm test}(E)$ are
      0.91, 0.85 kcal/mol. The MAE$_{\rm train}(F)$ and MAE$_{\rm
        test}(F)$ on forces for the model are 0.39, 0.75
      kcal/(mol$\cdot$\AA), and the corresponding RMSE$_{\rm
        train}(F)$ and RMSE$_{\rm test}(F)$ are 1.46, 3.54
      kcal/(mol$\cdot$\AA). }
    \label{fig:valid-tf-5162}
\end{figure}

\begin{table}[h!]
\centering
\caption{Comparison of energies for 5 stationary points between the
  predicted energies and {\it ab initio} energies (in kcal/mol) from
  the PhysNet base model.}
\begin{tabular}{|c|c|c|c|c|c|}
\hline
\textbf{} & H$_2$COO & TS-dioxirane & Dioxirane & TS-HCOOH & HCOOH \\
\hline
CCSD(T)      & 0.00  & 20.03        & -26.15    & 33.86    & 9.15  \\
\hline
base model   & 0.00  & 20.03        & -26.15    & 33.86    & 9.15  \\
\hline
\end{tabular}
\label{tab:base}
\end{table}

\newpage

\begin{table}[h!]
\hspace{2.0cm}Reactant \hspace{4.0cm}  Product
    \centering
 \begin{tabular}{c|c|c|c|c||c|c|c|c}
\hline
    \hline
    Bond harmonic & $k_b$ & $r_e$ & & & $k_b$ & $r_e$ & & \\
    \hline
    1 - 2 & 410.12 & 1.08 & & &473.13 & 1.09 &  \\
    1 - 3 & 410.12 & 1.08 & & &473.13 & 1.09 &  \\
    1 - 4 & 474.32 & 1.27 & & &437.81 & 1.38 &  \\
    4 - 5 & 203.53 & 1.34 & & & 242.86 & 1.52 &  \\
 \hline
    Bond Morse & $D_e$ & $r_e$ & $\beta$ & & $D$ & $r_e$ & $\beta$ & \\
    \hline
    1 - 5 & X & X & X &  & 127.75 & 1.38 & 2.69 \\
\hline
Angle & $k_\theta$ & $\theta_e$ &  & & $k_\theta$ & $\theta_e$ & & \\
\hline
2 - 1 - 3 & 26.39 & 121.99 & & & 20.13& 135.59 &   \\
2 - 1 - 4 & 46.27 & 116.96 & & & 54.50 & 118.82  &  \\
3 - 1 - 4 & 53.38 & 113.91 & & & 67.83 & 117.68 &  \\
1 - 4 - 5 & 39.38 & 121.72 & & & 67.78& 120.94 &  \\
4 - 1 - 5 & X & X & & & 22.97& 89.04 &  \\
2 - 1 - 5 & X & X & & & 82.66& 106.65 &  \\
3 - 1 - 5 & X & X & & & 54.73& 118.00 &  \\
1 - 5 - 4 & X & X & & & 40.18& 98.21 &  \\
\hline
Dihedral & N & $k_d$ & $\phi_d$  & & N & $k_d$ & $\phi_d$& \\
\hline
2 - 1 - 4 - 5 & 2 & 4.02 & 180.00 & & X & X  & X \\
3 - 1 - 4 - 5 & 2 & 5.54 & 180.00 & & X & X & X \\
\hline
Improper & N & $k_i$ & $\phi_i$  & & N & $k_i$ & $\phi_i$& \\
\hline
1 - 2 - 4 - 3 & 0 & 21.33 & 0.00 &  & 0 & 18.04 &0.00   \\
1 - 5 - 4 - 2 & 0 & X & X & & 0 & 15.26& 0.00  \\
\hline
GVDW & $r$ & $\epsilon$ & n & m & $r$ & $\epsilon$ & n  & m\\
\hline
1 - 5 & 1.76 & 0.72 & 6.19 & 12.82 & X & X  & X & X \\
\hline
\end{tabular}
   \caption{The harmonic bond, Morse bond, valence angle and generalized van
     der Waals parameters for reactant(CH$_2$OO) and
     product(dioxirane). $k_b$ in kcal/mol/\AA\/$^2$, $r_e$ in \AA\/,
     $D_e$ in kcal/mol, $\beta$ in \AA\/$^{-1}$, $k_\theta$ in
     kcal/mol/radian$^2$, $\theta_e$ in degree, $k_d$ in kcal/mol,
     $\phi_d$ in degree, $k_i$ in kcal/mol/radian$^2$, $\phi_i$ in
     degree, $r$ in \AA\/ and $\epsilon$ in kcal/mol. Atom
       number code (see Figure \ref{fig:timeseries-angle}): C(1), H$\rm
       _A$(2), H$\rm _B$(3), O$\rm _A$(4), O$\rm _B$(5).}
    \label{tab:dioxirane_channel}
\end{table}

\begin{table}[h!]
    \centering
    \begin{tabular}{c||c|c|c|c|c|c}
    
     $k$   & $V^0_{ij,k}$ & $\sigma_{ij,k}$ & $a_{ij,k0}$ & $a_{ij,k1}$ & $a_{ij,k2} $ & $a_{ij,k3} $\\
     \hline
     \hline
     1   & -5.36605 & 3.91184 & -2.86507 & -0.43938 & -0.20009 & 0.00187 \\
     2 & 13.43213 & 1.78751 & -0.21920 & 0.27335 & &  \\
     3 & 5.14916 &  3.94423 & -2.00850 & & & \\
    \end{tabular}
    \caption{GAPO parameters for dioxirane channel: $i$ labels the
      reactant, $j$ labels the product, $V^0_{ij,k}$ is the center of
      the Gaussian function (in kcal/mol), $\sigma_{ij,k}$ is the
      width of the Gaussian (in kcal/mol) and $a_{ij}$ is the
      polynomial coefficient in kcal/mol. }
    \label{tab:gapo-dioxirane}
\end{table}

\begin{table}[h!]
\hspace{2.0cm}Reactant \hspace{4.0cm} Product \centering
 \begin{tabular}{c|c|c|c|c||c|c|c|c}
\hline
    \hline
    Bond harmonic & $k_b$ & $r_e$ & & & $k_b$ & $r_e$ & & \\
    \hline
    1 - 3 & 413.90 & 1.09 & & &333.44 & 1.10 &  \\
    1 - 4 & 719.68 & 1.28 & & &388.47 & 1.23 &  \\
 \hline
    Bond Morse & $D$ & $r_e$ & $\beta$ & & $D$ & $r_e$ & $\beta$ & \\
    \hline
1 - 2 & 78.55 & 1.08 & 2.16 &  & X & X & X \\
5 - 2 & X & X & X &  & 67.21 & 0.96 & 1.98 \\   
4 - 5 & 349.82 & 1.34 & 1.02 &  & 9.83 & 1.53 & 1.00 \\
\hline
Angle & $k_\theta$ & $\theta_e$ &  & & $k_\theta$ & $\theta_e$ & & \\
\hline
2 - 1 - 3 & 18.18 & 106.37 & & &X  & X &   \\
2 - 1 - 4 & 50.22 & 115.51 & & & X& X&  \\
3 - 1 - 4 & 59.82 & 110.03 & & & 31.33 & 101.58 &  \\
1 - 4 - 5 & 99.54 & 120.17 & & & 92.80 & 111.12 &  \\
2 - 5 - 4 & X & X & & & 39.59 & 91.73 &  \\
\hline
Dihedral & N & $k_d$ & $\phi_d$  & & N & $k_d$ & $\phi_d$& \\
\hline
2 - 1 - 4 - 5 & 2 & 5.97 & 180.00 & & X & X  & X \\
3 - 1 - 4 - 5 & 2 & 7.37 & 180.00 & & X & X & X \\
3 - 1 - 4 - 5 & 3 & X & X & & 3 & 20.52  & 0.00 \\
1 - 4 - 5 - 2 & 2 & X & X & & 2 & 7.43 & 180.00 \\
\hline
Improper & N & $k_i$ & $\phi_i$  & & N & $k_i$ & $\phi_i$& \\
\hline
1 - 2 - 4 - 3 & 0 & 35.41 & 0.00 &  & X & X & X   \\
\hline
GVDW & $r$ & $\epsilon$ & n & m & $r$ & $\epsilon$ & n  & m\\
\hline
5 - 2 & 3.32 & 1.3 & 6.90 & 13.8 & X & X  & X & X \\
1 - 2 & X & X & X & X & 4.61 & 1.35  & 6.95 & 12.86 \\
\hline
\end{tabular}
 \caption{The harmonic bond, Morse bond, angle and generalized van der
   Waals parameters for reactant(CH$_2$OO) and product(HCOOH). $k_b$
   in kcal/mol/\AA\/$^2$, $r_e$ in \AA\/, $D$ in kcal/mol, $\beta$ in
   \AA\/$^{-1}$, $k_\theta$ in kcal/mol/radian$^2$, $\theta_e$ in
   degree, $k_d$ in kcal/mol, $\phi_d$ in degree, $k_i$ in
   kcal/mol/radian$^2$, $\phi_i$ in degree, $r$ in \AA\/ and
   $\epsilon$ in kcal/mol. Atom number code(see Figure
   \ref{fig:timeseries-angle}): C(1), H$\rm _A$(2), H$\rm _B$(3),
   O$\rm _A$(4), O$\rm _B$(5).}
    \label{tab:htransfer_channel}
\end{table}

\begin{table}[h!]
    \centering
    \begin{tabular}{c||c|c|c|c}
    
     $k$   & $V^0_{ij,k}$ & $\sigma_{ij,k}$ & $a_{ij,k0}$ & $a_{ij,k1}$  \\
     \hline
     \hline
     1  & -9.51488 & 28.24630 & -5.64736 & -0.11152 \\
     2 & -0.48168 & 6.31526  & -5.29915 &    \\
     3 & 34.98611 &  15.02948 & 2.19718 &  \\
    \end{tabular}
    \caption{GAPO parameters for H-transfer channel: $i$ labels the
      reactant, $j$ labels the product, $V^0_{ij,k}$ is the center of
      the Gaussian function (in kcal/mol), $\sigma_{ij,k}$ is the
      width of the Gaussian (in kcal/mol) and $a_{ij}$ is the
      polynomial coefficient in kcal/mol. }
    \label{tab:gapo-htransfer}
\end{table}

\end{document}